\documentclass[a4paper,fleqn,usenatbib]{mnras}

\usepackage{mathptmx}
\usepackage{pdflscape}
\usepackage[T1]{fontenc}
\usepackage{ae,aecompl}
\usepackage{hyperref}

\usepackage{wrapfig}
\usepackage{epstopdf}
\usepackage[latin1]{inputenc}

\usepackage{times}
\usepackage{mathptmx}
\usepackage{graphicx}	% Including figure files
\usepackage{amsmath}	% Advanced maths commands
\usepackage{amssymb}	% Extra maths symbols
\usepackage{xspace}
\usepackage{siunitx}
\usepackage{caption}
\usepackage[normalem]{ulem}
\usepackage{float}
\usepackage{CJK}
\usepackage{microtype}
\usepackage{booktabs}

\usepackage[version=4]{mhchem}

\newcommand{\NpHp}{N$_2$H$^+$}

\newcommand{\HCOp}{HCO{$^+$}}
\newcommand{\HTCOp}{H$^{13}$CO{$^+$}}
\newcommand{\Msun}{M$_\odot$}
\newcommand{\Lsun}{L$_\odot$}
\newcommand{\kms}{{km~s$^{-1}$}}
\newcommand{\unitMsunyr}{M$_\odot$ yr$^{-1}$}

\newcommand{\RN}[1]{
  \textup{\uppercase\expandafter{\romannumeral#1}}}      
\newcommand{\vin}{v$_{\rm in}$}

\newcommand{\vhill}{$v_{\rm in, Hill5}$}
\newcommand{\disp}{$\sigma_{v}$}
\newcommand{\dopb}{$\Delta v_b$}

\begin{document}
% Title of the paper, and the short title which is used in the headers.
% Keep the title short and informative.
\title[The infall profile in SDC335.579-0.292]{An inverted infall profile for the collapse of the massive star-forming IRDC SDC335.579-0.292}

% The list of authors, and the short list which is used in the headers.
% If you need two or more lines of authors, add an extra line using \newauthor
\author[Xie Jinjin et al.]{
{\begin{CJK*}{UTF8}{gkai}Xie Jinjin（谢津津）\end{CJK*}}$^{1, 2, 3, 4}$,
Gary~A. Fuller$^{2, 5, 6}$,
{\begin{CJK*}{UTF8}{bkai}Di Li（李菂）\end{CJK*}}{}$^{7, 3}$\thanks{E-mail: dili@mail.tsinghua.edu.cn, gary.a.fuller@manchester.ac.uk},
\newauthor Rowan Smith$^{8,2}$,
Nicolas Peretto$^{9}$,
Jingwen Wu$^{4}$,
{\begin{CJK*}{UTF8}{bkai} Yongxiong Wang（王永雄）\end{CJK*}}{}$^{2}$,
\newauthor {\begin{CJK*}{UTF8}{bkai}Yan Duan（段言）\end{CJK*}}{}$^{10, 3}$,
{\begin{CJK*}{UTF8}{gkai}Jifeng Xia (夏季风) \end{CJK*}}{}$^{3,4}$,
 Jarken Esimbek$^{1, 11, 4}$, Willem A. Baan$^{1,12}$
\\
% List of institutions
$^{1}$State Key Laboratory of Radio Astronomy and Technology, Xinjiang Astronomical Observatory, Chinese Academy of Sciences, 150 Science 1-Street, Urumqi, Xinjiang 830011, China\\
$^{2}$Jodrell Bank Centre for Astrophysics, Department of Physics \& Astronomy, The University of Manchester, Manchester M13 9PL, UK\\
$^{3}$National Astronomical Observatories, Chinese Academy of Sciences, A20 Datun Road, Chaoyang District, Beijing 100101, China\\
$^{4}$University of Chinese Academy of Sciences, Beijing 100049, China\\
$^{5}$Intituto de Astrof\'isica de Andalucia (CSIC), Glorieta de al Astronomia s/n E-18008, Granada, Spain\\
$^{6}$I. Physikalisches Institut, University of Cologne, Z\"ulpicher Str. 77, 50937 K\"oln, Germany\\
$^{7}$New Cornerstone Science Laboratory, Department of Astronomy, Tsinghua University, Beijing 100084, China\\
$^{8}$SUPA School of Physics and Astronomy, University of St Andrews, North Haugh, St Andrews, Fife, KY17 9SS, UK\\
$^{9}$School of Physics \& Astronomy, Cardiff University, Queen\textquotesingle s Building, The Parade, Cardiff, CF24 3AA, UK\\
$^{10}$Space Engineering University, Beijing 101416, China\\
$^{11}$ Xinjiang Key Laboratory of Radio Astrophysics, Urumqi 830011, China \\
$^{12}$Netherlands Institute for Radio Astronomy, ASTRON, 7991 PD Dwingeloo, The Netherlands\\
}

% These dates will be filled out by the publisher
\date{Accepted XXX. Received YYY; in original form ZZZ}

% Enter the current year, for the copyright statements etc.
\pubyear{2024}

% Don't change these lines

\label{firstpage}
\pagerange{\pageref{firstpage}--\pageref{lastpage}}

\maketitle

\begin{abstract}
There is increasing evidence for global collapse of clumps over parsec-scales in massive star formation regions. Such collapse may result in characteristic molecular line emission profiles but the spatial variation of such lines has rarely been quantitatively examined. Here we explore the infall properties using the spatially-resolved {\HCOp} J=1--0 and {\HTCOp} J=1--0 maps of the massive infrared dark cloud (IRDC) SDC335.579-0.292. We compare the observations with the analytical  Hill5 model and radiative transfer models. This  shows that the best-fit infall velocity towards the cloud centre to be well-constrained to $-0.6$ to $-1.6$\,{\kms} and the mass infall rate between a few $\times10^{-3}$ and $10^{-2}$\,\Msun yr$^{-1}$. The comparison also highlights some limitations of the Hill5 method. We demonstrate that the width of optically thin spectral lines, which are usually interpreted as resulting from turbulent motions, are in fact dominated by unresolved, ordered infall motions within the beam.
Our results suggest a complex collapse situation where
there is a minimum in the infall velocity at $\sim2\times10^{18}$ cm (0.7\,pc) with the infall velocity increasing at both smaller and larger radii.
The parsec-scale infall with an inverted velocity profile indicates that the accretion in this massive star-forming cloud should have intermediate scales, at which fragmentation or filament formation has to occur before material flows onto the cloud centre.

\end{abstract}


%%%%%%%%%%%%%%%%%%%%%%%%%%%%%%%%%%%%%%%%%%%%%%%%%%

%%%%%%%%%%%%%%%%% BODY OF PAPER %%%%%%%%%%%%%%%%%%

\section{Introduction}\label{sec:intro}

Gravitational collapse is the basic step in star formation \citep[e.g.][] {1977ApJ...214..488S,1977ApJ...214..725E,2017MNRAS.467.1313V} and the nature and evolution of this collapse are central in determining the masses of the stars, hence their destinies. When gas is infalling, differences in the excitation temperatures across the cloud can cause self-absorption features in optically thick molecular line profiles with the emission to the blue side of the absorption being of a higher intensity than the red. Meanwhile, the optically thinner species peak in the self-absorption dip of the optically thicker lines \citep[e.g.][]{1977ApJ...211..122S, 1986ApJ...309L..47W,1987A&A...186..280A}.

Such characteristic molecular line profiles have been observed towards both low- and high-mass star-forming regions \citep[e.g.][]{1993ApJ...404..232Z, 1995ApJ...449L..65M, 2003ApJ...592L..79W,2005A&A...442..949F, 2010ApJS..188..313W}. Infall properties such as the velocity profiles and radius of the infall region have been studied with various methods and have been further used to calculate crucial star-forming parameters including mass accretion rates and accretion timescales \citep[e.g.][]{2010A&A...517A..66L,2013A&A...549A...5R,2018ApJ...862...63C,2023ApJS..269...38X}. However, analyses of hydrodynamic simulations have shown that a prominent blue-asymmetric line profile may not be observable, even if there is infall \citep{2012ApJ...750...64S}.

Early infall studies mainly focused on low-mass nearby regions including starless cores with the exploration of different tracers \citep[e.g.][]{1993ApJ...404..232Z,1995ApJ...449L..65M,2001ApJS..136..703L}. In recent years, more observations have revealed global collapse appearing in massive star-forming regions \citep[e.g.][]{2003ApJ...592L..79W,2005A&A...442..949F,2010ApJS..188..313W,2013A&A...555A.112P,2019ApJ...870....5J,2023ApJ...955..154Y,2023ApJ...957...61H,2026ApJ...998..167J}. Semi-analytic models such as the `two-layer' model \citep{1996ApJ...465L.133M} and the `Hill' model \citep{2005ApJ...620..800D}, use simplifying assumptions on the physical properties of a region to derive an analytic form for the blue asymmetric line profile and so provide an estimate of the infall velocity. These methods have been widely used to infer the infall velocities from molecular line observations of star-forming regions \citep[e.g.][]{2010A&A...520A..49S,2013ApJ...764L..14Z,2016MNRAS.456.2681Q,2021RAA....21..208X}. One-dimensional models including radiative transfer (RT) (e.g., RATRAN \citep{2000A&A...362..697H}) have been applied to derive infall velocity by simulating infall profiles and comparing these with observations \citep[e.g.][]{2013A&A...555A.112P,2021ApJ...909..199O,2023ApJ...955..154Y}. To fully understand the collapse scenario of a cloud, RT modelling of the observations is needed.

Simulations on infall profiles have been carried out towards star-forming regions. Detailed modelling of a low-mass star-forming region B335 \citep{2015ApJ...814...22E} shows the infall is consistent with a free-fall velocity profile ($\propto r^{-1/2}$ where $r$ is the distance from the central source) and L483 is well modelled by a profile with \vin$\propto r^{-1}$ \citep{2019A&A...629A..29J}. The three-dimensional line RT code, LIME (LIne Modeling Engine) \citep{2010A&A...523A..25B} has been recently used to study $^{13}$CO infall velocities varying within the infall region towards B335, from which the models could reproduce high infall velocities on larger scales and maintain low velocities close to the protostar simultaneously when infall velocities deviated from free-fall velocities \citep{2023A&A...677A..62B}. 
Studies of the density profiles in simulations of cores also suggest that the infall velocities may be larger at larger radii \citep{2021MNRAS.502.4963G}, an `outside-in' collapse \citep{1977ApJ...214..725E}.
However, such modelling of infall velocity profiles for studying global collapse, especially simulations for observational molecular line profiles, towards massive star-forming regions are still lacking. 

\subsection{The massive-star-forming IRDC SDC335.579-0.292}\label{sec:introSDC}

\begin{figure} 
  \centering
   \includegraphics[width=\columnwidth]{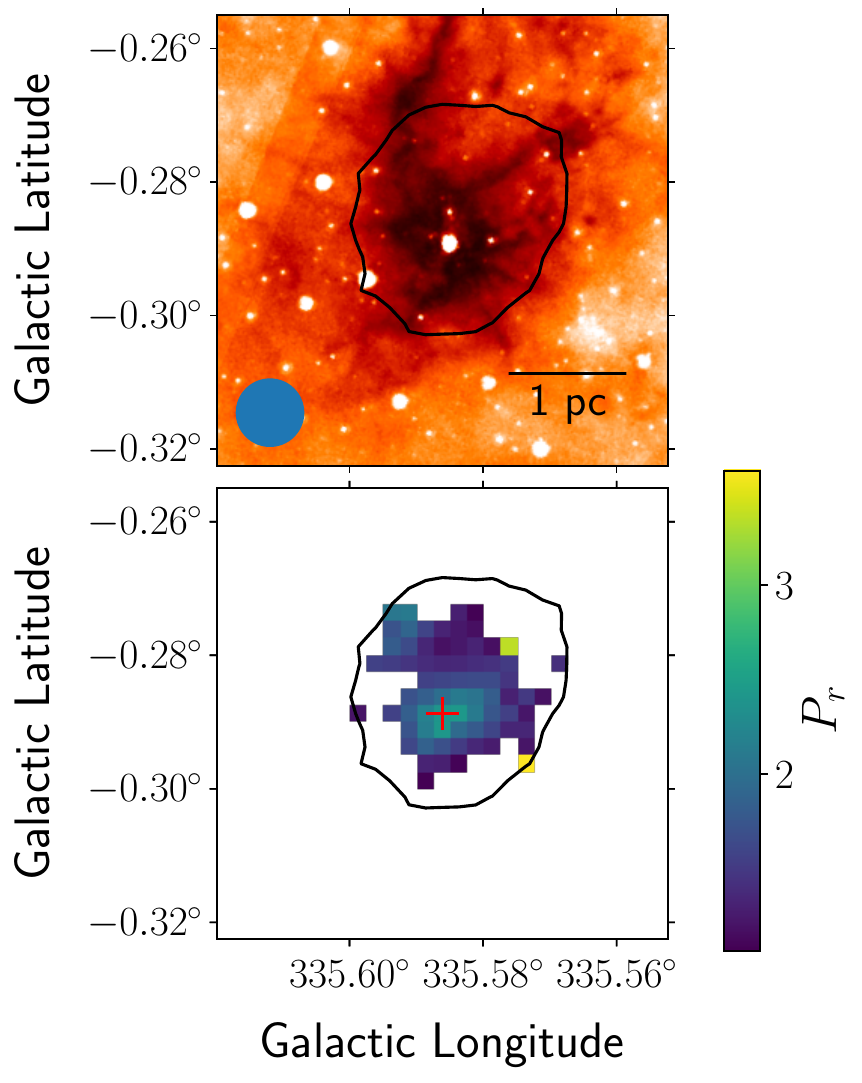}
\caption{{\it Upper:} Spitzer GLIMPSE \citep{2009PASP..121..213C} 8\,$\mu$m image of SDC335 showing the absorption tracing the densest regions of the clump.  The beam of the \ce{HCO+}\,J$=1-0$ observations is shown in the lower left and a 1\,pc scale bar is shown in the lower right. \textit{Lower:} The spatial distribution of the $P_r$, the intensity ratio of the blue-shifted peak to the red-shifted peak (Sec.~\ref{sec:linechara}). The parameter $P_r$ is evaluated for the region where the integrated intensity (the 0th-order moment emission map) is $>20$\% of the peak integrated intensity in the line (as indicted by the contour) on both panels. The red cross marks the pixel closest to the centre of the clump with the largest value of $P_r$. 
%The minimum value of $P_r$ shown is 1.06. 
}
\label{fig:infallParamMaps}
%Figure from notebook: AnalysisNotebooks/InfallFullCube.ipynb
\end{figure}

One extensively studied massive star-forming region is the Infrared Dark Cloud (IRDC) SDC335.579-0.292 (hereafter SDC335, also known as IRAS 16292-2722). SDC335 \citep[e.g.][]{2013A&A...555A.112P,2021A&A...645A.142A,2022ApJ...929...68O,2023MNRAS.520.3259X}, located at a distance of 3.25\,kpc, spans a diameter of 2.4\,pc \citep{2013A&A...555A.112P}. The total mass is calculated to be 5500 $\pm$ 800\,{\Msun} with 3 young OB stars forming in the region \citep{2013A&A...555A.112P,2015A&A...577A..30A}. Global collapse was identified in the HCO$^{+}$\,J$=1-0$ observations with infall signatures seen over the extended region of the clump \citep{2013A&A...555A.112P}. An overview of the region is shown in the upper panel of Figure~\ref{fig:infallParamMaps}.

To deduce the underlying velocity field that gives rise to a given line profile in a region like SDC335 is, however, a non-trivial task. Detailed RT simulations are needed to estimate infall velocities. \citet{2021A&A...645A.142A} showed that the outflow rates are consistent with mass accretion rates in the inner circumstellar region comparable to the mass accretion rate in the surrounding clump. Previously RATRAN \citep{2000A&A...362..697H} and the analytic `two-layer' model \citep{1996ApJ...465L.133M} were used to analyse the peak \HCOp\ spectrum of SDC335 and suggest that this cloud is collapsing at a speed of 0.7\,{\kms} \citep{2013A&A...555A.112P}. While simple to use and widely adopted to provide an estimate of infall velocities, parametric models like the `two-layer' model \citep{1996ApJ...465L.133M} and the `Hill' model \citep{2005ApJ...620..800D} (see Sec.~\ref{subsec:hill} for details)  assume a uniform infall velocity. However, it is unclear whether this is the case in collapsing regions. In this work, we carry out a modelling study of the spectral maps of SDC335 to spatially resolve the infall properties.

As noted by \citet{2013A&A...555A.112P}, SDC335 shows a spatially extended distribution of blue asymmetric \ce{HCO+}\,J$=1-0$ line profiles characteristic of infall. In Appendix~\ref{sec:linechara} we describe the possible parameterisation of this asymmetry. One such parameterisation is $P_r$, the ratio of the intensity of the blue-shifted peak to that of the red-shifted peak. 
Figure~\ref{fig:infallParamMaps} (lower panel)
shows the spatial distribution of $P_r$ across SDC335. 
The map shows that the spectroscopic signature of infall is seen over a significant portion of the clump.  
Note that at positions towards SDC335 (and in general) where the line profile is asymmetric (due to infall) but does not have clear double peak, the parameter $P_r$ is poorly defined. 
Moreover, while providing a description of the line shape, $P_r$ (and the other parameters described in Appendix~\ref{sec:linechara})  do not on their own provide any estimate of the properties of the infall. To do this requires a model. The simplest and most widely used are models based in a simple layer approximation \citep{1996ApJ...465L.133M}, with the most sophisticated of these being the `Hill' model \citep{2005ApJ...620..800D}.
For our analysis in this work we define the central position of the clump as the pixel close to the peak of the clump with the largest value of the parameter $P_r$. This pixel is at position ($l$,$b$)$=335.5836^\circ, -0.2862^\circ$ and is indicated by the red cross in Figure~\ref{fig:infallParamMaps}.

We describe the observational data and the models in Section~\ref{sec:data} and the methods we used to characterise the infall signatures and the quantitative comparison method to evaluate the results are described in Appendix~\ref{sec:linechara}. The analyses from the semi-analytic model are described in Section~\ref{sec:analytic}. RT models used are described in Section~\ref{sec:rt}, which also includes the results from RT models such as RATRAN \citep{2000A&A...362..697H}, RADMC-3D \footnote{\url{http://www.ita.uni-heidelberg.de/~dullemond/software/radmc-3d}} \citep{2012ascl.soft02015D}, and LIME \citep{2010A&A...523A..25B}, as well as the comparisons among the RT models and the semi-analytic model. The robustness of the semi-analytic model is examined through comparisons with the inputs of the RT models across the cloud. The effects from underlying assumptions such as uniform infall velocities on the results of models are explored and compared with the observed data. We discuss the discrepancies among the models, the sensitivity and the accuracy of the analytic model, and the implications on massive star formation in Section~\ref{sec:discussion}. Section~\ref{sec:concl} summarises our results which quantitatively reproduce the observations of SDC335 with our best-fit models and the obtained infall velocity and mass infall rate.

\section{Observational Data}
\label{sec:data}
We use the observations of {\HCOp}\,J$=1-0$ at 89.188518\,GHz \citep{1998JQSRT..60..883P} and 
{\HTCOp}\,J$=1-0$ at 86.7542884\,GHz together with 
N$_2$H$^+$\,J$=1-0$ at 93.173772\,GHz 
from the MALT90 survey undertaken with the Mopra 22\,m single-dish telescope\footnote{https://www.narrabri.atnf.csiro.au/mopra/}\citep{2011ApJS..197...25F,2013PASA...30...38F,2013PASA...30...57J}. The systemic velocity of the cloud, $-46.5$\,\kms, was derived from the optically thin lines H$^{13}$CO$^{+}$\,J$=1-0$ and {\NpHp}\,J$=1-0$. At this frequency, the angular resolution of the Mopra telescope is $\sim$37'' \citep{2005PASA...22...62L}, corresponding to 0.43~pc at the distance of SDC335. The {\HCOp}\,J$=1-0$ line has the blue asymmetric profile (Figure~\ref{fig:hill5_central}) characteristic of infall over a $\sim$3 pc-sized region of the clump \citep{2013A&A...555A.112P}. The velocity resolution is 0.1\,km s$^{-1}$
with the rms $\sim$ 0.2\,K on T$_\mathrm{mb}$ scale. We smoothed the data by 2.5 channels (giving a final spectral resolution of 0.25\,km s$^{-1}$), which reduced the rms to 0.013\,K as this allows better characterisation of the main features of the blue asymmetric profiles. In the following analysis, the models are all rebinned to match this resolution. 

The \NpHp\ observations are only used to provide an estimate of the dense gas velocity for comparison with the \HCOp\ spectra at some positions in the clump. The velocity at these positions is determined from a fit to the hyperfine components using the pyspeckit fitting routine adopting a frequency of 93.1737637\,GHz for the 
$F_1=2$, $F=3$ to $F_1=1$, $F=2$ hyperfine component \citep{2009A&A...494..719P}.

To account for the telescope efficiency and the coupling of the source to the telescope beam (the beam filling factor), when comparing the models with the observations, the peak intensities of the modelled spectral lines were scaled by a factor $\eta$ to match the peak intensity of the observed line.
\citet{2005PASA...22...62L} gives efficiencies for the Mopra telescope of $\eta_{mb}=0.49$ for compact sources and 0.65 for more extended sources such as the case for SDC335. 
% This isn't actually true....
The scaling factor applied to the models, $\eta$, was always $<0.7$, which is consistent with the measured telescope efficiency within the uncertainties. 
So the emission from models that would require an efficiency significantly larger than $\sim0.7$ would be too weak to match the observations. 
On the other hand, values of $\eta<0.7$, would correspond to the source not completely filling in the telescope beam. In the modelling discussed below the values of $\eta$ required to match the models to the observations were consistent with the range $0.49<\eta< 0.7$, except for the RATRAN model, which required the slightly larger value of 0.8.

%We will discuss the effect of smooth in ~\ref{subsec:Hill_accuracy}.
% GAF: Added this as showing the extend infall signature is important.

\section{Infall Properties: Semi-analytic Model}\label{sec:analytic}

%First we use the commonly used Hill model to estimate properties of the infall. While Hill is simple to use and widely used to provide an estimate of infall velocities, but assumes a uniform infall velocity in the infalling region. 

To probe the infall properties in SDC335, we first use 
a one-dimensional semi-analytic model, the `Hill' model. The `Hill' model is considered to be an upgraded model of `two-layer' model \citep{1996ApJ...465L.133M}, which tends to underestimate the infall velocities \citep{2005ApJ...620..800D}. The `Hill' model assumes a hill-shape excitation temperature distribution of gas across the cloud, as compared with the earlier `two-layer' model which adopted two excitation temperature layers in the cloud. Hill5 is a model with five free parameters, which are optical depth $\tau$, infall velocity $v_{in}$, systematic velocity $v_{\rm lsr}$, velocity dispersion $\sigma$, and peak excitation temperature $T_{p}$. The Hill5 has been considered the most robust model among the Hill models with different sets of free parameters \citep{2005ApJ...620..800D}. The Hill5 model has been applied to calculate infall velocities in other massive star-forming regions including other IRDCs, as well as low mass star-forming regions \citep[e.g.][]{2006ApJ...636..952W,2018ApJ...861...14C,2018ApJ...862...63C}.

\subsection{Hill5 Model Description}
\label{subsec:hill}

%To analyse the infall profiles we first use the Hill, one-dimensional analytic model \citep{2005ApJ...620..800D}.
In the `Hill' model, the brightness temperature of the molecular line above the background, $\Delta T_{B}\left(v\right)$, is given by
\begin{eqnarray}
\begin{aligned}
  \Delta T_{B}\left(v\right)&=&\left(J\left(T_{P}\right)-J\left(T_{0}\right)\right)
  \left[\left(1-e^{-\tau_{f}\left(v\right)}\right) \Big/ \tau_{f}\left(v\right)\right.\\
&&\left.-e^{-\tau_{f}\left(v\right)}\left(1-e^{-\tau_{r}\left(v\right)}\right)\Big/\tau_{r}\left(v\right)\right] \\
&&+ \left(J\left(T_{0}\right)-J\left(T_{b}\right)\right)\left[1-e^{-\tau_{r}\left(v\right)-\tau_{f}\left(v\right)}\right],
\end{aligned}
\label{eq:Hill}
\end{eqnarray}
where $T_{B}$ is the brightness temperature defined as $T_{B} = \left(c^{2}\Big/2{\nu}^{2}k\right)I_{v}$, and $I_{\nu}$ is the specific intensity. $J(T)$ refers to the Planck average energy or the Planck-corrected brightness temperature where $J\left(T\right) = \left(h\nu\Big/k\right)\left[\exp\left(h\nu\Big/kT\right)-1\right]^{-1}$. The excitation temperature at the near and far edges of the source, along the line of sight, is $T_{0}$. The optical depth of the foreground material where the excitation temperature is rising along the line of sight is $\tau_{f}$, while $\tau_{r}$ is the optical depth along the line of sight in the region beyond the peak excitation temperature.  A uniform velocity dispersion $\sigma$ is assumed for the entire cloud. The optical depths at velocity $v$ for the foreground and background (rear) layers are given by 
\begin{eqnarray}
\begin{aligned}
  \tau_{f}\left(v\right)&=&\tau\exp\left[-\left(v-v_{lsr}-v_{\rm in,Hill5}\right)^{2}/2\sigma^{2}\right],
\end{aligned}
\label{eq:tauf}
\end{eqnarray}
\begin{eqnarray}
\begin{aligned}
  \tau_{r}\left(v\right)&=&\tau\exp\left[-\left(v-v_{lsr}+v_{\rm in,Hill5}\right)^{2}/2\sigma^{2}\right].
\end{aligned}
\label{eq:taur}
\end{eqnarray}

A Python fitting code using Python package LMFIT \footnote{https://lmfit.github.io/lmfit-py/} was developed with the above equations, which requires an initial guess as an input for the five free parameters, $\tau$, infall velocity $v_{\rm in,Hill5}$\footnote{We adopt the convention that the infall velocity, \vin, is negative for motion towards the centre of the region. In comparisons of the speed of the infall we will use the magnitude of \vin, such that faster infall will correspond to more negative values, and this is referred to as larger values throughout the text.}, systematic velocity $v_{lsr}$, velocity dispersion $\sigma$, and peak excitation temperature T$_{p}$\footnote{Equation~\ref{eq:Hill} is typeset incorrectly in the Astrophysical Journal in (\citet{2005ApJ...620..800D}. The correct version can be found in the arXiv version of the paper.}. We also used the Hill5 fitter in the PySpecKit \citep{2011ascl.soft09001G}  package\footnote{https://pyspeckit.readthedocs.io/en/latest/hill5infall\_model.html\#module-pyspeckit.spectrum.models.hill5infall} and found to be in excellent agreement with our own independent implementation of the routine which was developed to confirm the form of the equations.

%\textcolor{blue}{Added the equations above. Please check them. }\textcolor{red}{OK, but not all of these variables are in the equation above. Need to add the equations which show what these variables are.} 
%The initial guess will not affect the fitting results from the model as long as an error estimate can be given. An error estimate can only be given when the 3$\sigma$ confidence interval can be reached. 

 %We use it here to characterise both the observed line profiles and those produced by the RT models.

\subsection{The Central Pixel} 

\begin{figure} 
  \centering
  \includegraphics[width=\columnwidth]{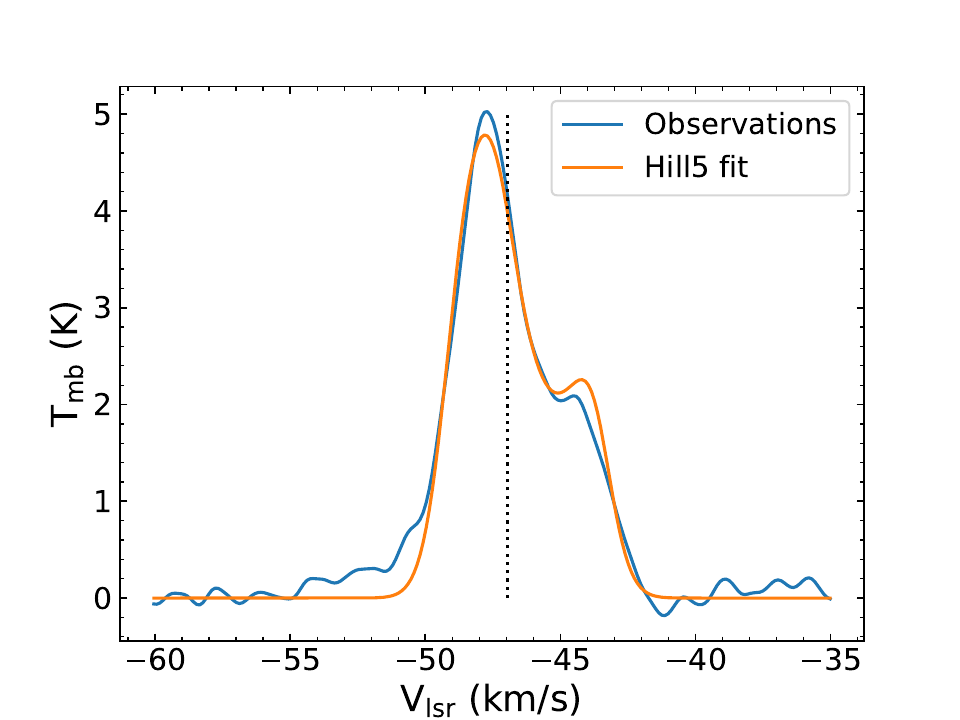}
\caption{Spectrum towards the central pixel and the Hill5 fit to it. The fitted parameters are listed in Table~\ref{tab:bestFitHill5Models}. The dotted vertical line indicates the central velocity of N$_2$H$^+$ J=1--0, $-46.9\pm0.1$\,\kms. This velocity is somewhat different from the velocity derived from the Hill5 fit which refers to the central velocity of the currently infalling, lower density gas.
}
\label{fig:hill5_central}
\end{figure}

\begin{figure*} 
   \centering
   \includegraphics[width=0.8\textwidth]{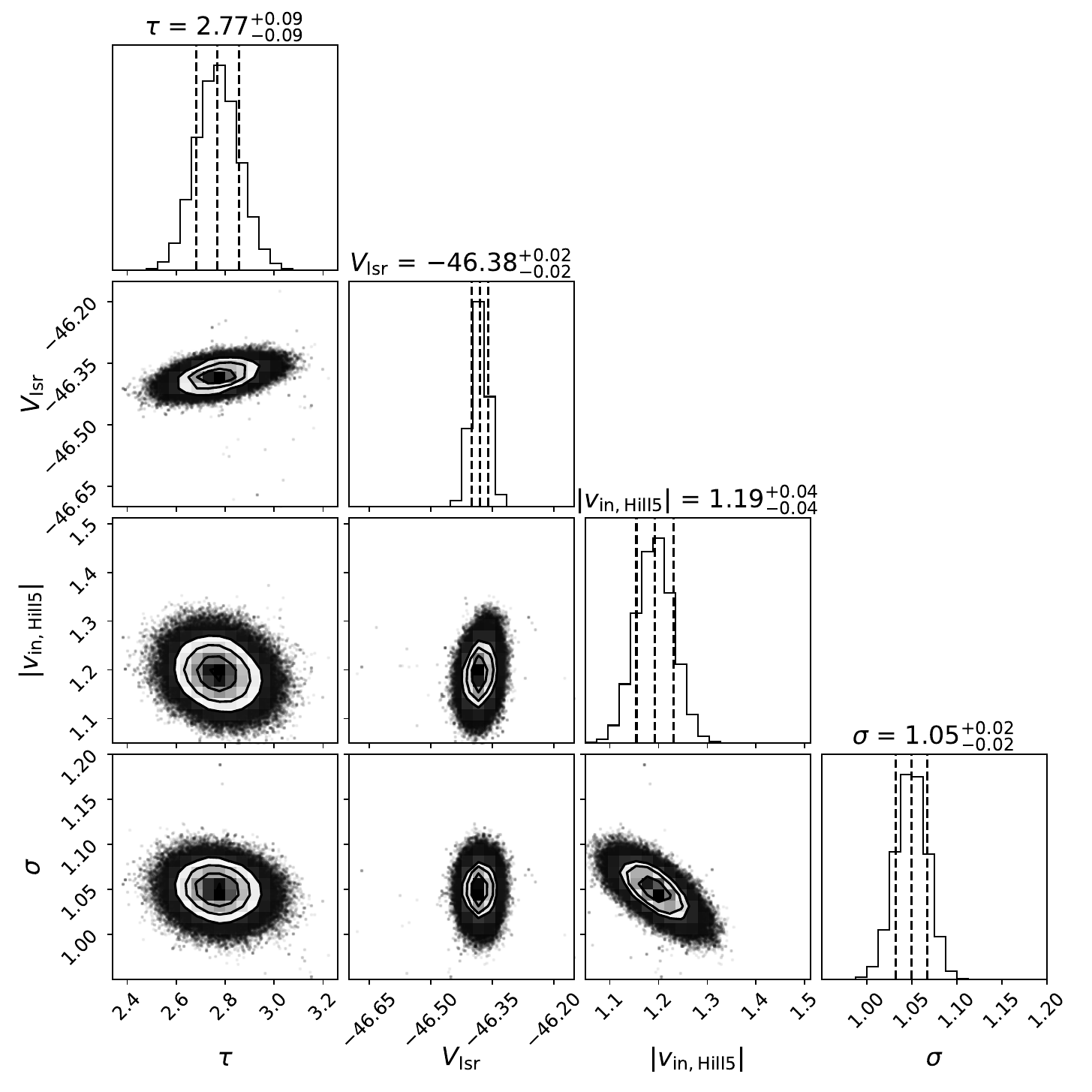}
\caption{Corner plot showing the distribution of fit variables for 50 000 Markov Chain Monte Carlo (MCMC) evaluations of the Hill5 fit to the central \HCOp\,J$=1-0$ spectrum. The marginal distributions for each variable shown at the top of each column are labelled by the median value and their 1-$\sigma$ equivalent uncertainties. The lower panels show the distribution of the fit values in the parameter planes. The distribution of points in high density parts is shown in contours and as points in the lower density regions. The peak excitation temperature was included as part of fit, but is not shown here. }
\label{fig:mcmcCorner}% From: ./Python/Hill5FitMap.ipynb
\end{figure*}

%The Hill5 fit to the central pixel of the observed \ce{HCO+} J=1-0 map is shown in Figure~\ref{fig:hill5_central}. 
The best-fit results of infall parameters for the central pixel on the observed \ce{HCO+}\,J$=1-0$ map from Hill5 are listed in Table~\ref{tab:bestFitHill5Models} and the fitted spectra are shown in Figure~\ref{fig:hill5_central}. The uncertainties are estimated from  Markov Chain Monte Carlo (MCMC) sampling used with $5\times10^5$ samples. Figure~\ref{fig:mcmcCorner} shows the distribution of the fit parameters 
with their mean and effective 1-$\sigma$ uncertainties (as given by the 16$^{\mathrm{th}}$ and 84$^{\mathrm{th}}$ percentiles of the distributions). The marginalised distributions show that there is some correlation between the fitted $v_{lsr}$ and the optical depth ($\tau$). A correlation is also seen between the velocity dispersion ($\sigma$) and infall velocity (v$_{\rm in,Hill5}$), although both are well constrained with formal uncertainties of $\pm0.02$ \kms\ and $\pm0.04$ \kms, respectively. 

%{Describe results and the method. Mention fit is shown in figure 12.  }

%%% Table of parameters of best fit HILL5 model

\begin{table}
\caption{Best-fitting Hill5 parameters for the central pixel.}
\label{tab:bestFitHill5Models}
\centering
\setlength{\tabcolsep}{3pt} % tighten horizontal padding
\renewcommand{\arraystretch}{1.05}
\begin{tabular}{c c c c c}
\hline
$\tau$ & $v_{\rm lsr}$ & \vhill & \disp & $T_p$ \\
 & (\kms) & (\kms) & (\kms) & (K) \\
\hline
$2.77\pm0.09$ & $-46.38\pm0.02$ & $-1.19\pm0.04$ & $1.05\pm0.02$ & $10.71\pm0.20$ \\
\hline
\end{tabular}
\end{table}

% From: ./Python/FinalPlots.ipynb

\subsection{The Spatial Distribution of the Infall Velocity}

\begin{figure} 
   \centering
   % Figure from Notebook: Hill5DispMap
   \includegraphics[width=0.45\textwidth]{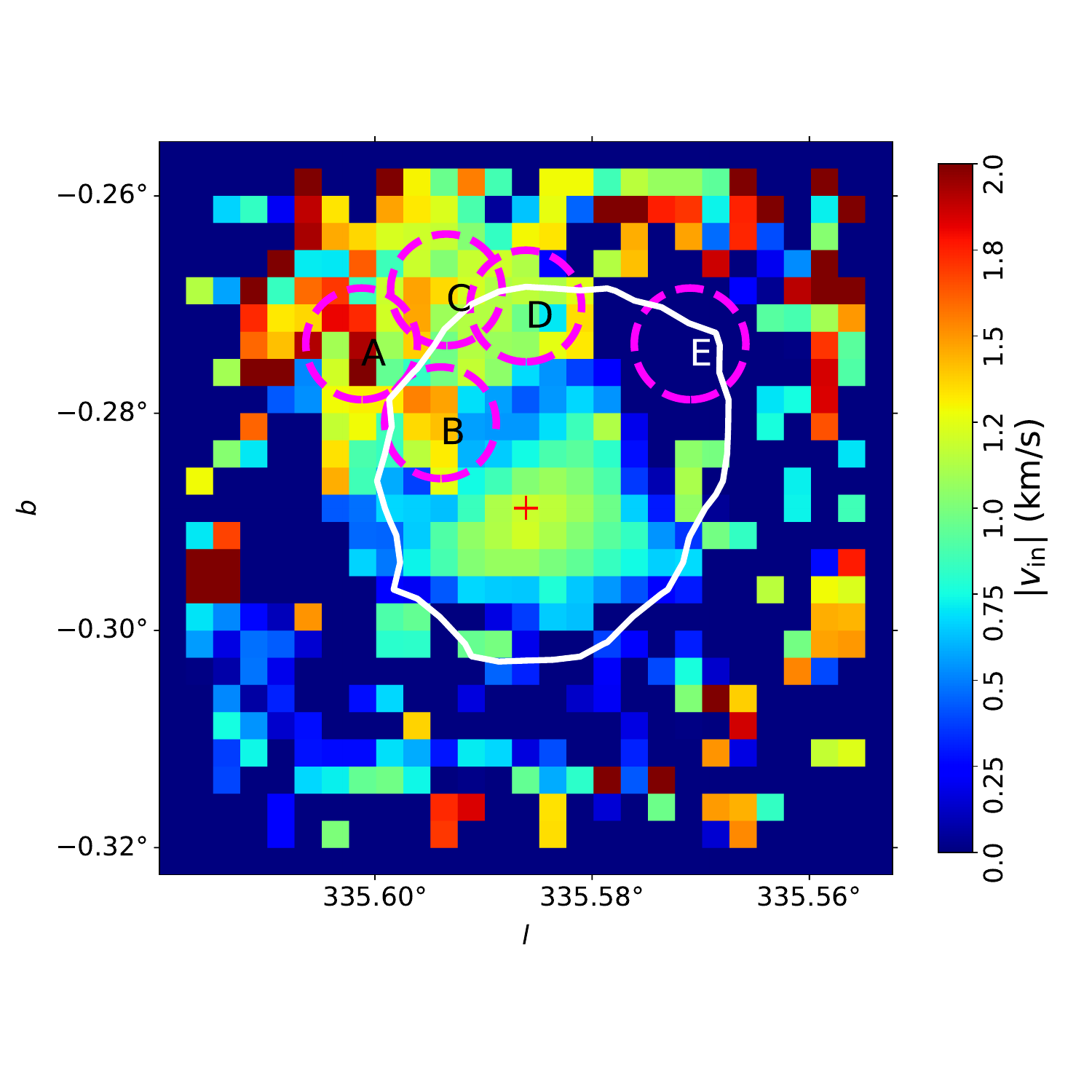}
\caption{The spatial distribution of the magnitude of the infall velocity ($|v_{\textrm in}|$) derived from Hill5. The red cross marks the central pixel. The magenta circles show the regions over which the spectra shown in Figure~\ref{fig:offsetSpec} are extracted. The white contour labels of the 20\% of the peak integrated intensity.}
\label{fig:hill5Map}
\end{figure}

The map of the infall velocities estimated by Hill5 for all the spectra in the \HCOp\ cube is shown in Figure~\ref{fig:hill5Map}. The map shows that moving outward from the central pixel, the fitted infall velocity initially decreases. Over most of the map, the infall velocity reaches close to zero around the contour corresponding to the line reaching 20\% of its peak flux. However, towards the northeast, the Hill5 fit gives consistently larger values for v$_{\rm in,Hill5}$ than the $[-0.75,-0.5]\,\mathrm{km\,s^{-1}}$ in the minimum around the central pixel, but the values have significant scatter, pixel to pixel.  
%{Describe what is seen in the figure}

\begin{figure} 
   \centering
   \includegraphics[width=\columnwidth]{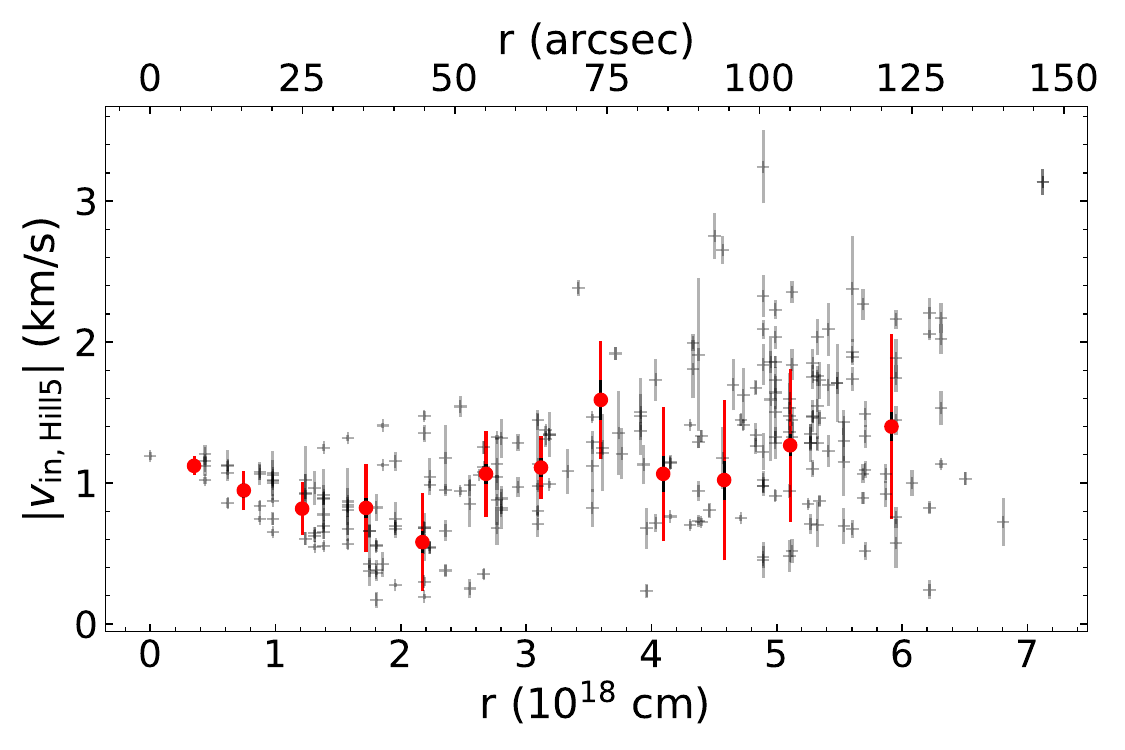}
\caption{Hill5-fitted infall velocity, \vhill, as a function of radius from the central pixel. The grey points show the values for the individual pixels in the map with the uncertainties from their fits which are typically $\sim0.1 - 0.2$\,\kms. The red points show these binned in bins of 10''. The red points are plotted at the average radius of the points averaged in that bin. The red error bars show the dispersion of values used to calculate the average.
% GAF {Add fits to annular averages as shown in later version of plot. Figure \ref{annspec5}?}
}
% figure from Notebook: Hill5DispMap
\label{fig:hill5Rad}
\end{figure}

Figure ~\ref{fig:hill5Rad} shows the infall velocity variation as a function of the radius from the central pixel from the map in Figure~\ref{fig:hill5Map}.
%{towards a particular direction?}, 
 The red points show the infall velocity averaged over radial bins 10'' wide. Although there are significant uncertainties, both the values for individual pixels and the binned values show the derived infall velocity decreases in magnitude from the central pixel out to a radius of about 40''($\sim2\times10^{18}$ cm or 0.65\,pc).
 Beyond this, there is a significant scatter (as seen in Figure~\ref{fig:hill5Map}), though the binned values suggest a relatively uniform value of \vin\ of $\sim-1$ \kms.

\begin{table}
    \centering
        \caption{Sizes of the annular regions used to construct averaged spectra.}
    \label{tab:annuli}
    \begin{tabular}{cccccc}
    \hline
    Aperture     & Inner radius &    Outer radius & \multicolumn{2}{c}{Aperture central radius} \\
    &  (arcsec) & (arcsec) & (arcsec)  & ($10^{18}$cm)\\
    \hline
    A1 & 18.6 &  37.2 & 28 & 1.4\\ 
    A2 & 37.2 &   55.8  &47 & 2.3\\
    A3 & 55.8 &  73.8  &65 & 3.2\\
    \hline
    \end{tabular}
    % Defined in notebook AnnAverSpec
%    galactic; annulus(335.586105, -0.288735, 0.31' 0.62')
%    galactic; annulus(335.586105, -0.288735, 0.62' 0.93')
%    galactic; annulus(335.586105, -0.288735, 0.93' 1.23')
\end{table}

%\begin{table}
%\centering
%\caption{Best fit parameters to the central pixel for models with a uniform \vin, a density profile with $p=1.5$ (except where noted) and a uniform temperature of 20 K. (Note $\sigma_v = \Delta v_b/\sqrt{2}.$)}	
%\label{tab:bestfit_central_pixel}
%%\vspace{-1mm}\footnotesize
%\begin{tabular}{lccc} 
%\hline
%Models & \vin ({\kms}) & \disp ({\kms}) &
%\hline
%%RATRAN & 0.8 & 0.8 & 6.6\\
%% Updated 2 Feb 2021 - Re-checked models. 
%
%%T = 20 K \\
%%\cline{1-1}
%%RATRAN & 0.8 & 1.0 & 7.2 \\
%%LIME p=1.5 & 1.3 & 1.06 & 7.7\\
%% LIME p=2 & 1.3 & 1.06 & 8.9 \\
%%RADMC-3D & 1.6 & 1.3 & 8.5\\
%%HILL5 & 1.2 & 1.05 & -  \\
%%T Profile\\
%%\cline{1-1}
%%LIME & 1.1 & 1.1 & 7.4 \\
%\hline
%\end{tabular}
%\end{table}
%
%%% Table of parameters of best fit models

\begin{table*}
  \begin{center}
  \caption{Best fit parameters to the central pixel from the RT models. } \label{tab:bestFitModels}

    \begin{tabular}{cclcccccc}
    \hline
   & & & &  &  &  Peak \\
 Label& $p$ & Model & $v_c$ & \dopb & \disp &  Intensity & $\eta$ & $R$\\

       &     &  & (\kms) & (\kms) &  (\kms) & (K) & & \\
     \hline
    LIME\\
           \cmidrule{1-1}
  & 1.5\\
$M_{LIME,0}$  & &  $f=1/2$ & -1.40 & 1.50 & 1.06 & 7.9 & 0.64 & 1.11 \\ 
$M_{LIME,1}$    & & $f=1$ & -1.10 & 1.60 & 1.13 & 7.5 & 0.67 & 1.09\\
  $M_{LIME,2}$ &  &$f=1$, T=20\,K & -1.40 &1.40  & 1.06 & 8.0 & 0.63 & 1.06 \\
       % Model above is from Grid/Grid5e and is the smaller v_in of two
      % points with the same quality of fit, the other has v_in= -1.4,
      % Dv 1.4 
 $M_{LIME,3}$ &  &          $f=3/2$ & -1.00 & 1.60 & 1.13 & 7.7 & 0.65 & 1.08\\
   & 2.0\\
  $M_{LIME,4}$ &   & $f=1/3$ & -1.40 & 1.50 &  1.06  & 8.9 & 0.57 &  1.13\\%
   $M_{LIME,5}$ &   & $f=1/2$ & -1.50 & 1.50 & 1.06 & 9.3 & 0.54 & 1.13\\%
 $M_{LIME,6}$ &      & $f=1$ & -1.10 & 1.60 & 1.13 & 8.7 & 0.57 & 1.11\\
 $M_{LIME,7}$   &      & $f=1$, T=20\,K  & -1.30 & 1.50  & 1.06 & 8.8 & 0.57& 1.05  \\
        % Model above is from Grid/Grid6e-p_2-20
 $M_{LIME,8}$   &     & $f=1$, 2R$^*$ & -1.30 &  1.60& 1.13 & 9.2 & 0.55 & 1.10 \\
        % Model in VConst_p2_Tmin20_2xsize	
    $M_{LIME,9}$  &     & $f=3/2$ & -0.80 & 1.70 & 1.20 & 8.0 &  0.63 & 1.12\\
  $M_{LIME,10}$   &      & $f=3$ & -0.60 & 1.70 & 1.20 & 8.1 & 0.62 & 1.11\\
   $M_{LIME, 11}$   &     & \vin$\propto r^{-0.5}$ & -1.50 & 1.50 & 1.06 & 9.0 & 0.56 & 1.13  \\
      \\
 
       RADMC-3D\\
       \cmidrule{1-1}

$M_{RADMC3D}$   &  1.5     & $f=1$, T=20\,K& -1.60 & 1.83 & 1.30 & 8.5 & 0.58 & 1.09\\
\\
RATRAN\\
       \cmidrule{1-1}
 $M_{RATRAN}$   &  1.5     & $f=1$,  T=20\,K & -0.90 & 1.60 & 1.13 & 6.3 & 0.78 & 1.16 \\
 % New RATRAN grid August 2024
 %        & $f=1$, T=20\,K & -0.9 & 1.60 & & 6.3 & 0.78 & \\
 \hline
        \end{tabular}
             \vspace{2ex}\\
  {\raggedright  Note: All the models have a temperature profile as described in the text,    except those noted as T=20\,K which have a uniform temperature of 20\,K. $^*$Model with outer radius
    twice that of the other models. The uncertainties on both $v_c$ and \dopb\ are estimated to be $\sim0.1$\,\kms, the sampling used in generating the grids of simulations. The peak intensities listed are not corrected for the telescope efficiency. However, an efficiency factor has been applied when model spectra are plotted for comparison with the observations. Note that  $\sigma_v=\Delta v_b/\sqrt{2}$.
 \par}
% 13 March 2026: Used line peak T of 5.03 K to calculate 

    \end{center}
  \end{table*}
  %% Updated 13 Sept 2024
  %% The best fit models are all linked in
  %% TProfiles/SmallModels/BestFits

\begin{figure} 
   \centering
%   \includegraphics[width=\columnwidth]{placeHolder.pdf}
% Figure from notebook Hill5ObsAnnFit
      \includegraphics[width=1.\columnwidth]{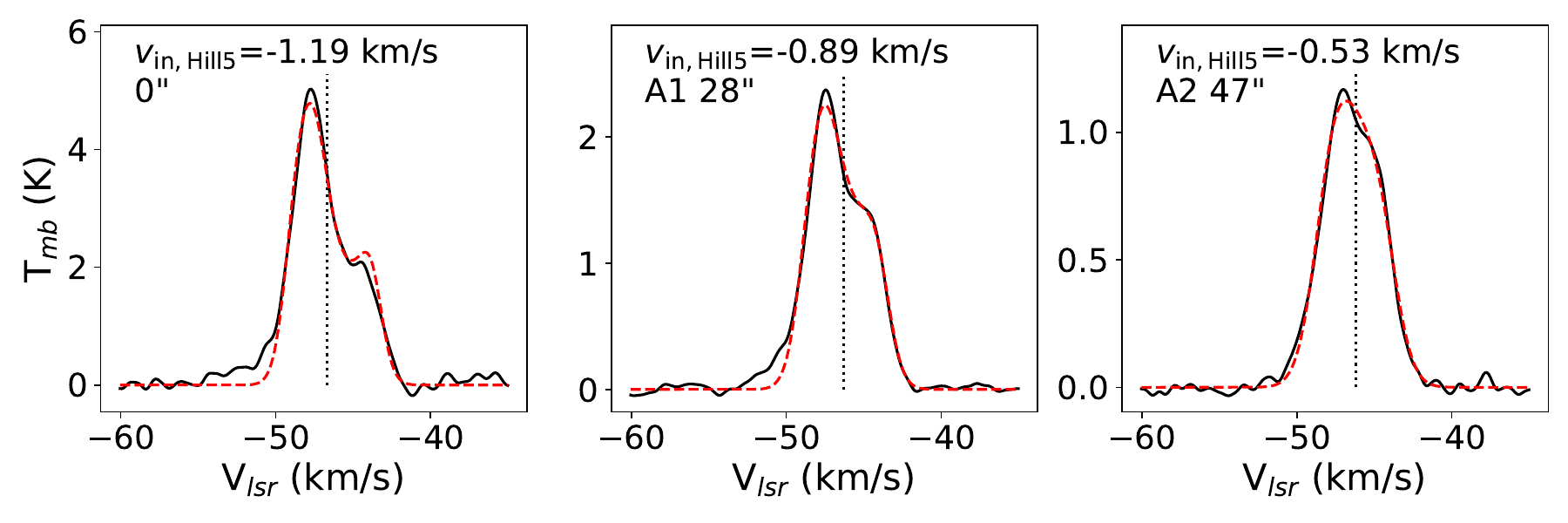}
\caption{Annular average spectra and Hill5 fits. The panels show the \HCOp spectra at the central pixel (left) and averaged over annuli of width 18.6$\arcsec$ with radii of 28$\arcsec$ (middle) and 47$\arcsec$ (right) respectively (as described in Table~\ref{tab:annuli}). The black curve shows the data and the vertical markers show the velocity derived from the \NpHp\ spectra averaged over the same regions. The dashed red curve shows the Hill5 fits to the spectra with the fitted infall velocity indicated in the respective panel. 
% GAF {Estimate infall velocities using mcmc ??}. 
}
\label{annspec5}
% Figures from notebook Hill5ObsAnnFit
\end{figure}

Hill5 has been widely applied to infer infall velocities and sometimes on the averaged spectra of a certain radius \citep[e.g.][]{2018ApJ...861...14C,2023ApJS..269...38X,2023ApJ...955..154Y}. For SDC335, we show the spectra and the fitted spectra in the central region averaged over annular regions in Figure~\ref{annspec5}, of which the details are listed in Table~\ref{tab:annuli}. The values for the infall velocity for these annularly averaged spectra are in excellent agreement with the averaged fit values in the comparable range of radius. They show the infall velocity increasing from $-1.19\pm0.02$\,\kms\ at the central pixel to $-0.58\pm0.01$\,\kms\  at a radius of 47'' ($2.2\times10^{18}$\,cm or 0.71\,pc). Hill5 therefore seems to consistently indicate a decrease in the infall velocity with increasing distance from the centre of the source to out to radii $\sim50$\arcsec ($\sim2.5\times10^{18}$\,cm or 0.81\,pc).

% GAF {any peculiar thing at 47"?}

\begin{figure*} 
   \centering
\includegraphics[width=0.41\columnwidth]{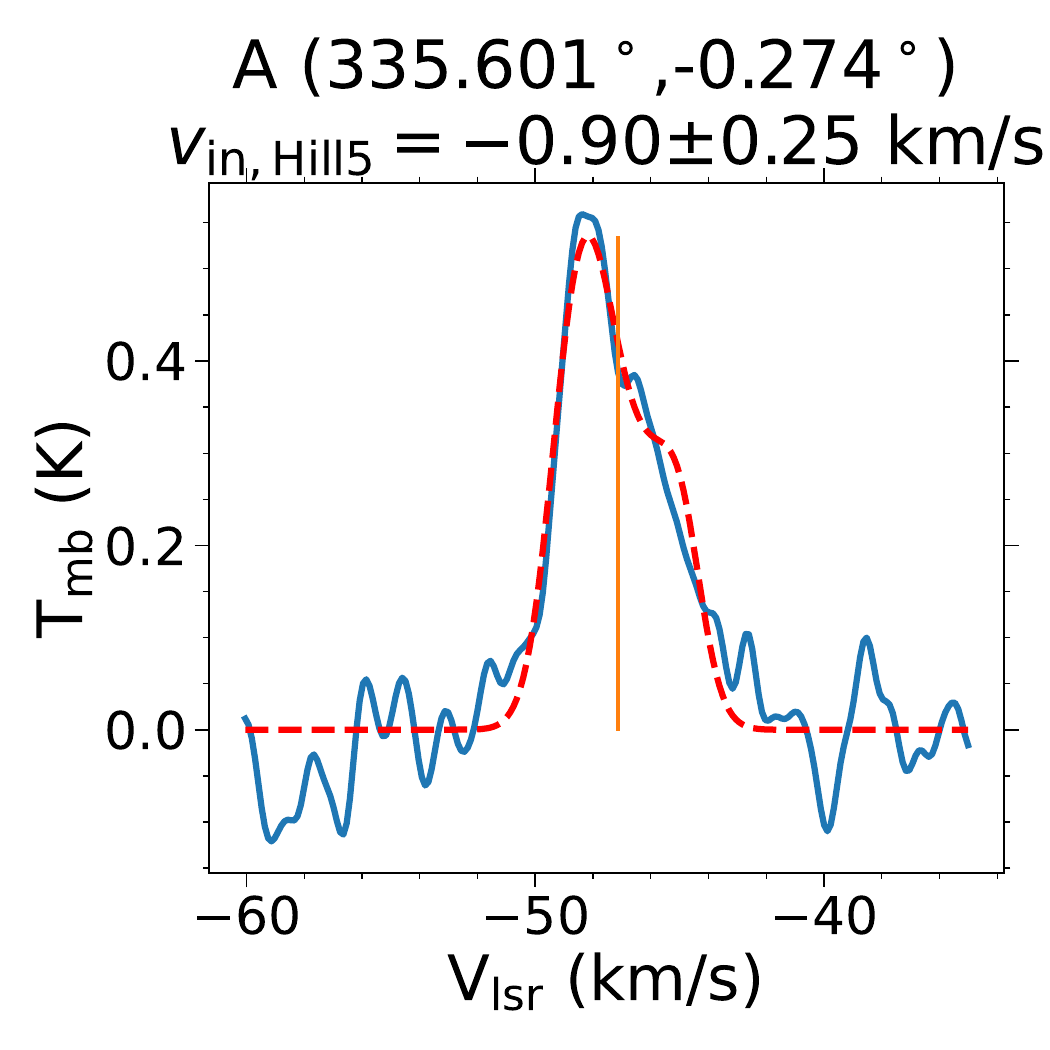}
\includegraphics[width=0.41\columnwidth]{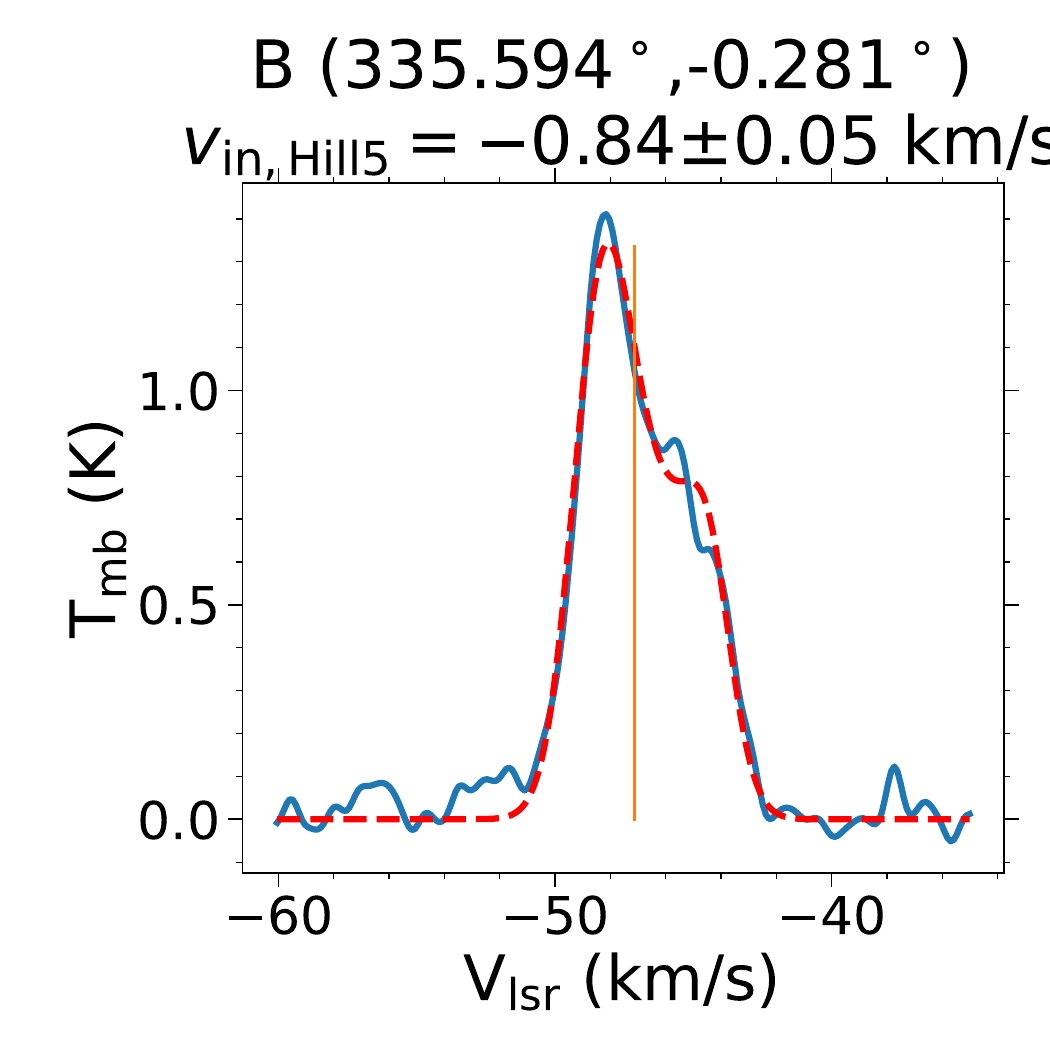}
\includegraphics[width=0.41\columnwidth]{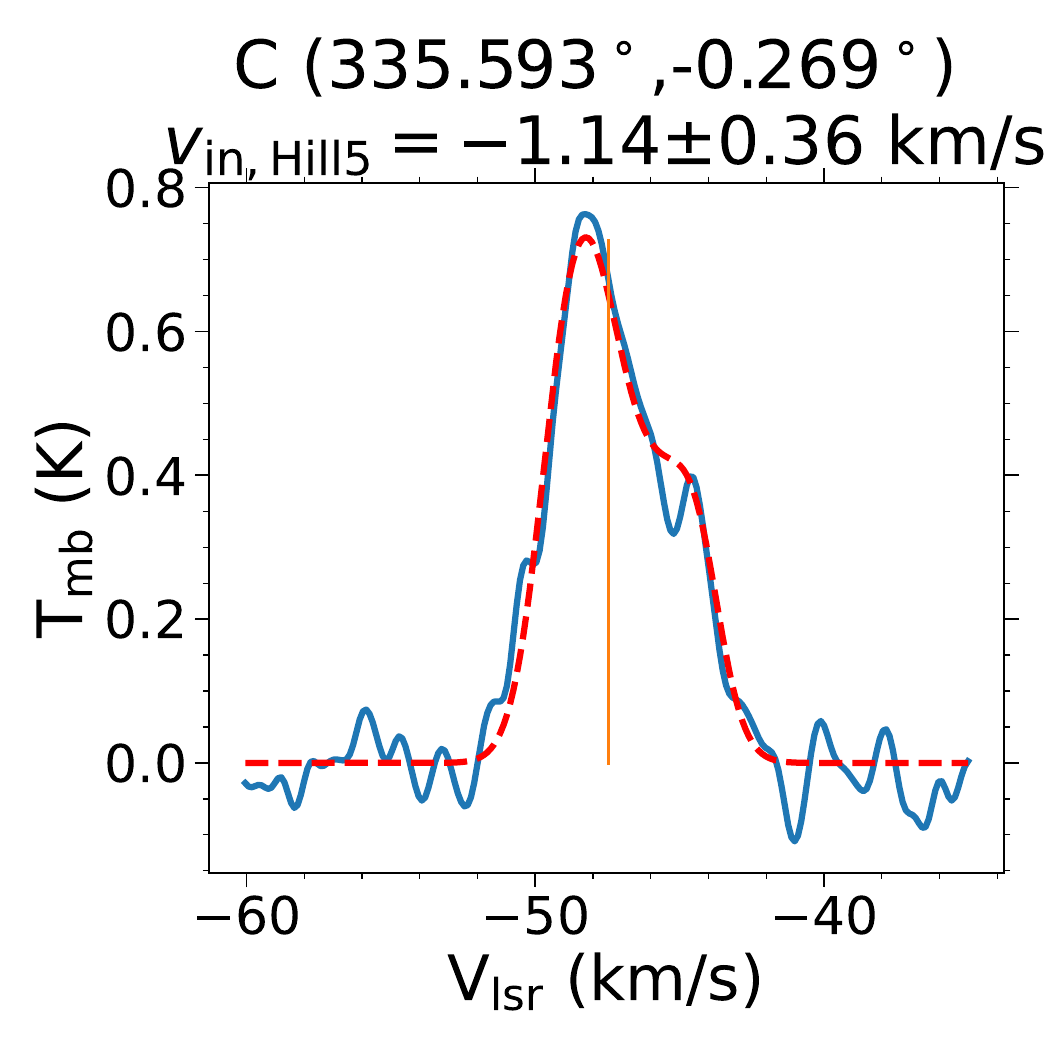}
\includegraphics[width=0.41\columnwidth]{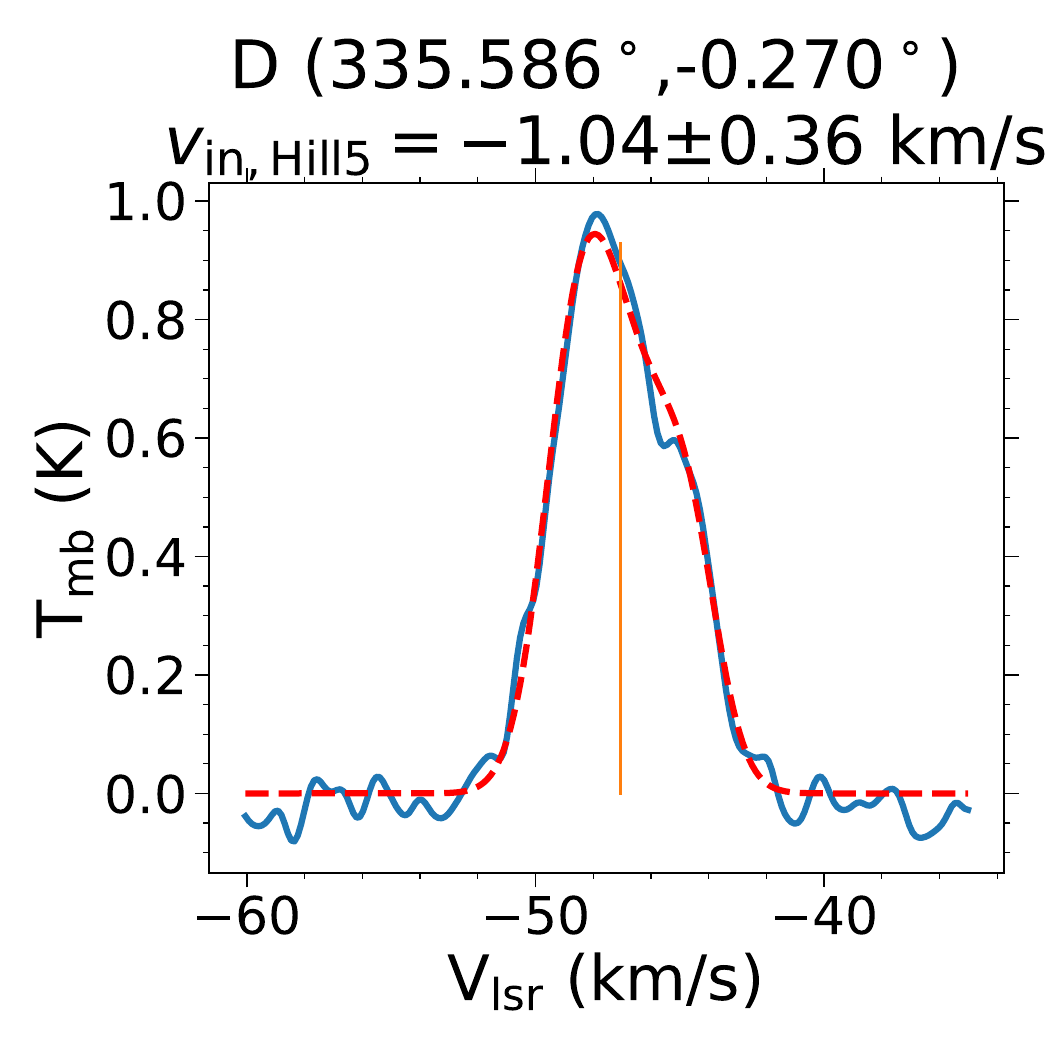}  
\includegraphics[width=0.41\columnwidth]{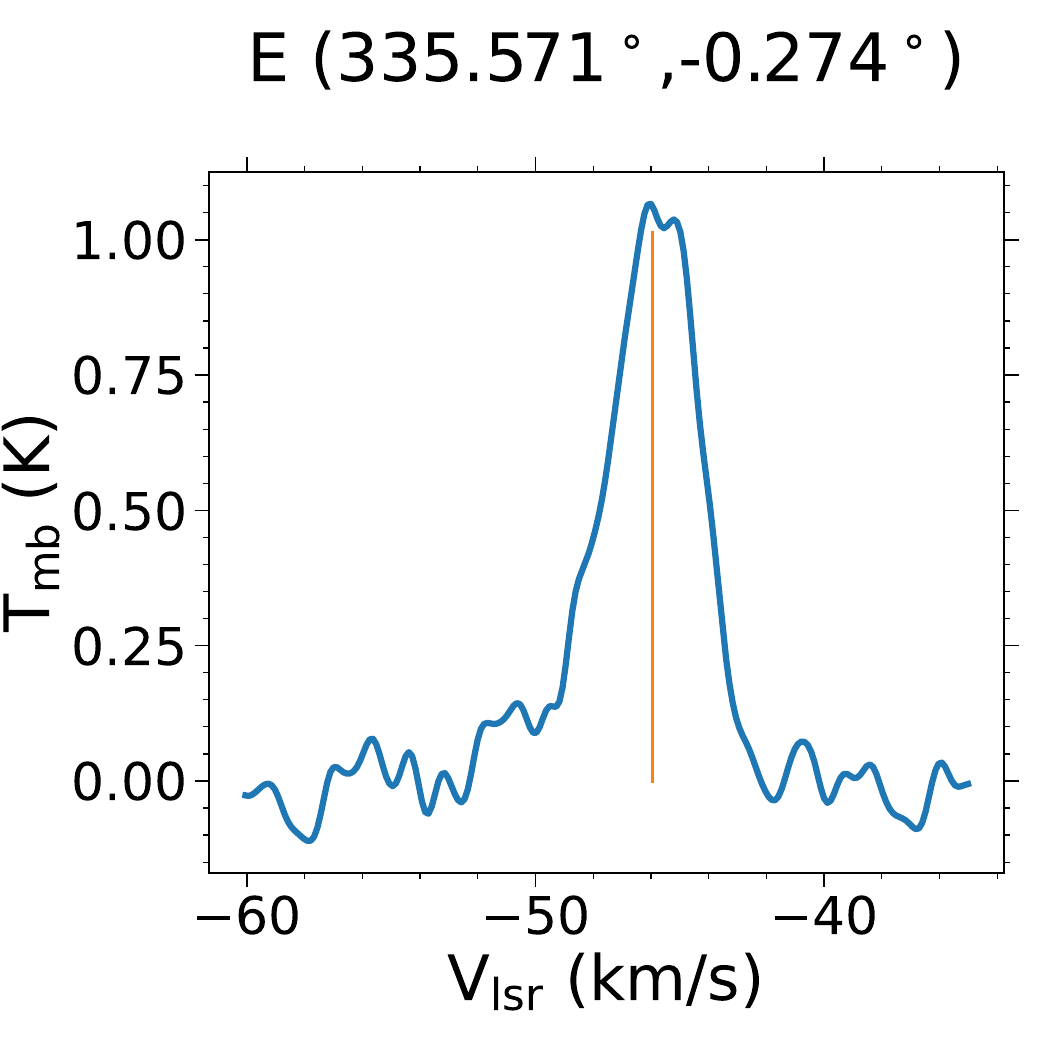}    

\caption{Spectra averaged over the circular regions shown in Figure~\ref{fig:hill5Map}.  
The central position (and label) of each spectrum is given  above each panel.
The data are shown in black and in the first four spectra the Hill5 fit is shown as the red dashed curve. For these positions the best fit infall velocity is shown in the title above the panel. Hill5 could not fit the spectrum at position E due to 
the lack of significant asymmetry in the line profile. The vertical line in each panel indicates the central velocity determined from a fit to the hyperfine structure of the \NpHp\ emission averaged over the same region as the \HCOp. }
\label{fig:offsetSpec}
% Plots from AnnAverSpec.ipynb
\end{figure*}

To explore the line profiles more distant from the centre, the data have been averaged over the five beam-sized regions shown in Figure~\ref{fig:hill5Map}. Four of these beam-sized regions were selected to sample the outer region of the clump where Hill5 found large infall velocities in the pixel-by-pixel fitting. The fifth region was selected as a test in a region where the pixel-by-pixel Hill5 fits found no infall.   
%{
The spectra and their Hill5 fits are shown in Figure~\ref{fig:offsetSpec}.
%} {
This figure shows that the four regions (labelled A to D) all show \HCOp\ spectra which are blue asymmetric (asymmetric with brighter blue peaks than red peaks)
%}  
with respect to the velocity of dense gas in the same region (as traced by \NpHp). The Hill5 fits indicate infall velocities in the range --0.8 to --1.1 \kms\ , consistent with the average values from fits to the individual spectra.
At these larger distances, $>50$\arcsec, Hill5 suggests an {\it increased magnitude} for the infall velocity towards the outer region compared to the region closest to the central source, over at least part of the clump. Region E in Figure~\ref{fig:hill5Map} covers an area where Hill5 could not identify the presence of infall and the aperture averaged spectrum of this region (Figure~\ref{fig:offsetSpec}), although not fully symmetric, does not show a blue asymmetry and the peak velocity agrees closely with that of \NpHp\ and so shows no evidence of infall. This is consistent with the larger Pr values found in the same region in Figure~\ref{fig:infallParamMaps}.

\section{Infall Properties: RT Models}\label{sec:rt}

%The collapse could also contribute to the velocity dispersion with internal turbulence \citep{2021A&A...655A...3H,2019A&A...625A..82N}.

To explore how the values derived by Hill5 map on to the actual infall properties in more complex situations, we have modelled clumps with a wider range of infall properties such as infall velocities and infall radii and compared the infall values estimated by Hill5 with the true infall properties.
%{Needs editing to focus on LIME but explain use of RadTran  etc. }
For this we adopted three RT models, namely LIME, RADMC-3D, and RATRAN. 
%To predict the line profiles from spherically symmetric physical models, we primarily used the LIME radiative transfer model. 
LIME is a 3-D RT non-LTE (Local Thermodynamic Equilibrium) accelerated Lambda Iteration code with uses unstructured 3-D Delaunay grids for photon transport  \citep{2010A&A...523A..25B}. RADMC-3D is a 3-D RT package  developed by \citet{2012ascl.soft02015D} for dust and molecular lines. The line transfer can be performed assuming LTE or the LVG (Large Velocity Gradient) approximation. RATRAN is a 1-D Cartesian grid Monte Carlo code which includes elements of accelerated Lambda iteration \citep{2000A&A...362..697H}.
% The RATRAN model in  \citet{2013A&A...555A.112P} was spherically symmetric and consisted of 20 concentric shells to simulate the emission.
 The molecule data files are from the Leiden database LAMDA \footnote{\url{https://home.strw.leidenuniv.nl/~moldata/}} \citep{2005A&A...432..369S}. 
 
 Initially, the comparison with the observations was focused on comparing the central pixel in the model  with the observed spectrum at the centre of the clump. Before this comparison, the models were convolved with a 37'' FWHM (full width at half maximum) Gaussian to simulate the Mopra beam at the frequency of \ce{HCO+}\,J$=1-0$ \citep{2005PASA...22...62L} and regridded to the same spectral channels as the observations.

\subsection{Physical Parameters for Models}
\label{seq:physmod}

To model the line profiles, we need to define a physical description of the properties of SDC335. Specifically these are the density profile, the temperature profile, turbulence, the abundance of \ce{HCO+}, as well as the infall velocity. Since the SDC335 clump appears round with little evidence of significant elongation at the resolution of the observations used here, we model SDC335 as a sphere of radius 1.2\,pc with the physical parameters dependent only on the radius, $r$, within the clump.

Modelling of high mass clumps has shown that their density distributions are well fitted with  density profiles of the form $\rho(r)\propto r^{-p}$ typically with $1\le p \le 2$
\citep[e.g.][]{2002ApJ...566..945B,2005A&A...434..257W,2014ApJ...785...42P,2016A&A...585A.149W}. The measurements of SDC335 show that it is relatively symmetric, highly centrally condensed, with \citet{2013A&A...555A.112P} showing that SDC335 has a mass distribution consistent with $p\sim2$. %For the modelling here 
We, therefore, adopted $p=2$ but also explored some models with $p=1.5$ for comparison with \citet{2013A&A...555A.112P} and to explore the sensitivity of our results to the assumed density profile. The normalisation of the density profile was constrained to match the observed total mass within 1.2\,pc of 5500 M$_\odot$ \citep{2013A&A...555A.112P}.

\begin{figure} 
   \centering
   \includegraphics[width=\columnwidth]{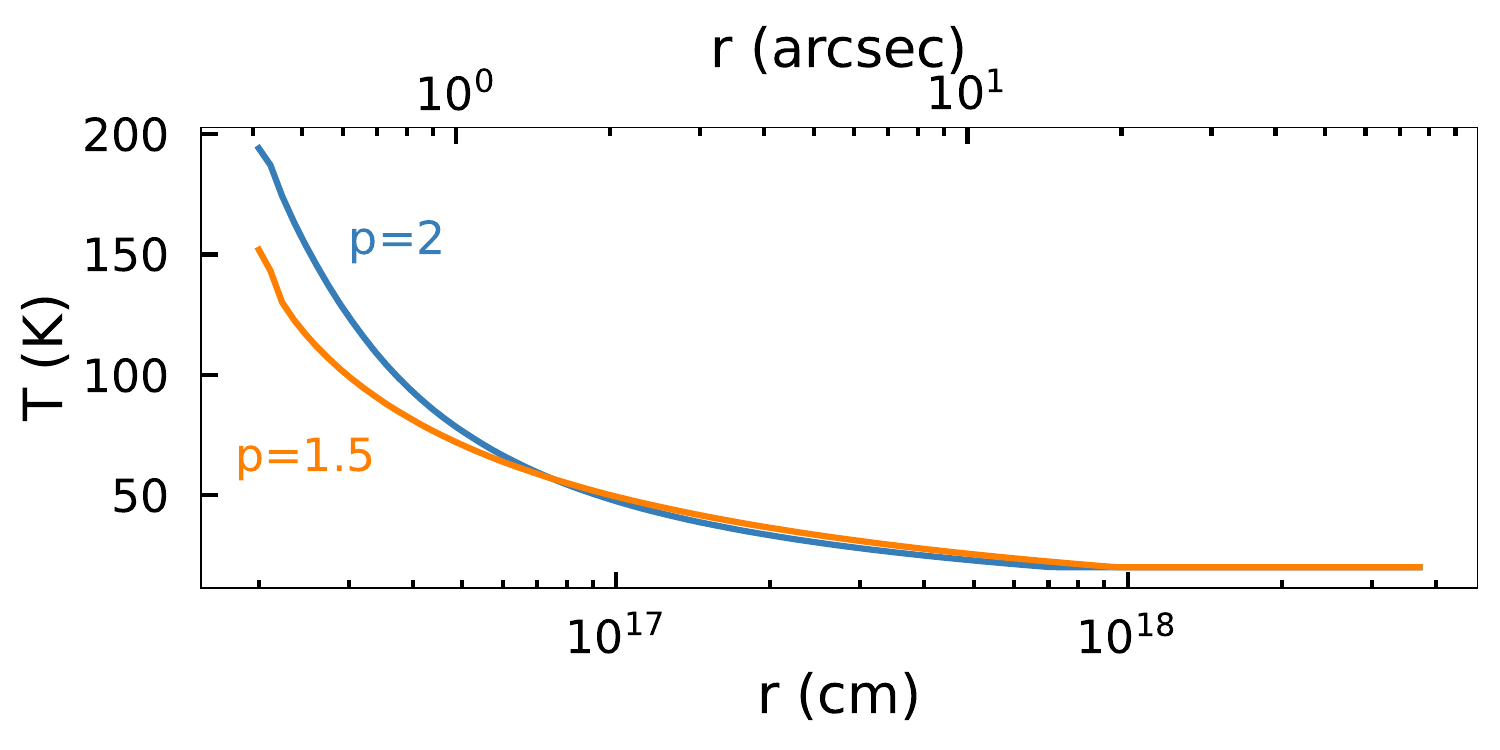}
\caption{Temperature profiles determined by RADMC-3D  for a central source of temperature 3000\,K and a total luminosity of $2\times10^4$\,\Lsun. The curves show the profiles for density profiles with $p=1.5$ and $p=2$ as labelled. The minimum temperature for both profiles was fixed at 20\,K.}
\label{fig:tempProfiles}
%
% Notebook: PlotLIMEParameters-Tprofile
\end{figure}

To determine the temperature structure for our spectral line models, we %{
used RADMC-3D, which can self-consistently calculate the temperature profile, to derive the temperature of our adopted density profiles
%} 
($p=2$ and $p=1.5$) assuming a central heating source. However, in the outer regions of the clump, the temperature was fixed at a minimum of 20\,K, consistent with the uniform temperature adopted by \citet{2013A&A...555A.112P} and the measured temperatures of the outer regions of IRDCs \citep[e.g.][] {2006ApJS..166..567R,2021SCPMA..6479511X}. The outflows from the embedded sources in SDC335 indicate a mass accretion rate onto the protostars of $1.4\times10^{-3}\,\mathrm{M_\odot\,yr^{-1}}$\citep{2021A&A...645A.142A}. At this high accretion rate 
\citet{2009ApJ...691..823H} and \citet{2010ApJ...721..478H} show %{
that the central protostar is bloated with a lower surface temperature than would be expected for its luminosity.
%} {
Therefore, in the RADMC-3D model, 
%{
we assumed the central source had a luminosity $2\times10^4$\,\Lsun\,\,and a surface temperature of 3000 K. 
%}
The resulting adopted temperature profiles for a $p=2$ and a $p=1.5$ density profile are shown in Figure~\ref{fig:tempProfiles}. In addition, some models were also run with a uniform temperature of 20\,K to allow both a more direct comparison with \citet{2013A&A...555A.112P} and also to assess the impact of the temperature profile on the line profiles. 

%The interior temperature is consistent with the protostar in SDC335 observed with ALMA \citep{2022ApJ...929...68O}. 

% ngVLA does not yet exist.!!! This paper is about a possible application. 
% The rapid accretion rate of massive protostars results in a bloated phase in the formation, which recently has been captured by ngVLA \citep{2021arXiv210308859T}. 

%This temperature structure shown in Figure \ref{fig:tempProfiles} is consistent what depicted by the bloated massive star formation phase, in which the interior temperature increases as the protostar contracts and a protostar accretes in a large radius to maintain a low effective temperature and low UV luminosity \citep{2010ApJ...721..478H,2013ApJ...778..178H,2016ApJ...824..119H}.

%{temperature profile in RADMC3D}\textcolor{blue}{Measured luminosity - bloated protostars - adopted parameters. }

We adopt a uniform HCO$^+$ abundance of $7\times10^{-10}$, consistent with the value used by \citet{2013A&A...555A.112P}.  As discussed by \citet{2013A&A...555A.112P}, at the galactocentric radius of SDC335 the expected abundance ratio \ce{^{12}C}/\ce{^{13}C} (and so the abundance ratio \ce{H^{12}CO+}/\ce{H^{13}CO+}) is 30, but \citet{2013A&A...555A.112P} measured a value of 20. We use this observed ratio in modelling the \ce{H^{13}CO+} lines. Also an abundance ratio of 30 results in lines that are too weak to match the observations.

%\subsubsection{Radial Infall Velocity Structure}
%\label{sec:velstructure}

We explore models where the infall velocity, $v_{in}(r)$, is a function of radius, $r$, within the clump. The form of 
%{
$v_{in}(r)$ adopted is
%} 
\begin{eqnarray}
v_{in}(r) &=&  
\begin{cases}
v_ c & r\le r_c\\
v_ c + v_ c(f-1) \frac{r-r_c}{R_0-r_c} & r_c<r\leq R_{0}\\
\end{cases}
\label{eqn:vprofile}
\end{eqnarray}
where $r$ is the distance from the clump centre, $r_c$ is the radius of the central region which has a uniform infall velocity of $v_{\rm c}$, $R_0$ is the maximum radius (1.2 pc), and the factor $f$ which describes the infall velocity at $R_0$ as a multiple of $v_{\rm c}$. A value of $f=1$ corresponds to a uniform infall velocity across the region. For all the models $r_c=0.3$\,pc which corresponds to the beam size of the \HCOp\ observations. 
The range of factors $f$ by which \vin\ increases and decreases with radius is selected based on an initial set of comparisons which suggested that values $1/3 \leq f \leq 3$ provided reasonable fits to the spatial distribution of the infall line profiles. In addition, a model with $p=2$ and where 
%{
the infall velocity is described by
%}
\begin{eqnarray}
v_{in}(r) &=&  
\begin{cases}
v_c & r\le r_c\\
v_c \left(\frac{r_c}{r}\right)^{1/2} & r_c< r \leq R_{0}, 
\end{cases}
\label{eqn:vprofileff}
\end{eqnarray}
to mimic a gravitational free-fall collapse is also investigated.

\subsection{The Central Pixel}

%To probe the range of infall properties of SDC335, we  explored models with a spatially uniform infall velocity across the clump,
%and a constant temperature, which are 
%similar to those in \citet{2013A&A...555A.112P}, and  models where the infall velocity varied with radius in the clump (Sec.~\ref{sec:velstructure}). 

%\input{table_bestfit_central_pixel.tex}

All the best fit models for the central pixel are summarised in Table~\ref{tab:bestFitModels}.
To assess the discrepancy between our modelled line profiles and the data we use the statistic, $R$, 
\begin{eqnarray}
% Rcsq
R&=& \frac{1}{N_c} \sum_i A_i\\
\text{where} && \\
A_i &=& 
\begin{cases}
{d_i}/{m_i}     &  \text{if } d_i\geq m_i\\
 {m_i}/{d_i}     &  \text{if } d_i<m_i,
\end{cases}
\end{eqnarray}
%% AMD
%\begin{eqnarray}
%A&=& \Sigma_i \frac{|d_i-m_i|}{d_i}
%\end{eqnarray}
$d_i$ and $m_i$ 
are the intensity of the observed spectrum and the model spectrum in pixel $i$ respectively, so the statistic measures the ratio to the model to the data. 
The statistic is calculated over the velocity range $w_{1/2}$ (Appendix~\ref{sec:linechara}) and $N_c$ is the number of spectral channels in this velocity range. The $R$ statistic is more suitable for comparing models and observations than the parameters often used to characterise observed line profiles (Appendix~\ref{sec:linechara}) since $R$ is well defined even if the line profiles do not have a clear double peak structure (Sec.~\ref{sec:introSDC}).
The statistic $R$ is also more suitable for fitting the overall line profile that the usual $\chi^2$ statistic which can result in fits which are dominated by differences between the model and observations in strong features at the expense of fitting weaker features.

\begin{figure} 
   \centering
   \includegraphics[width=\columnwidth]{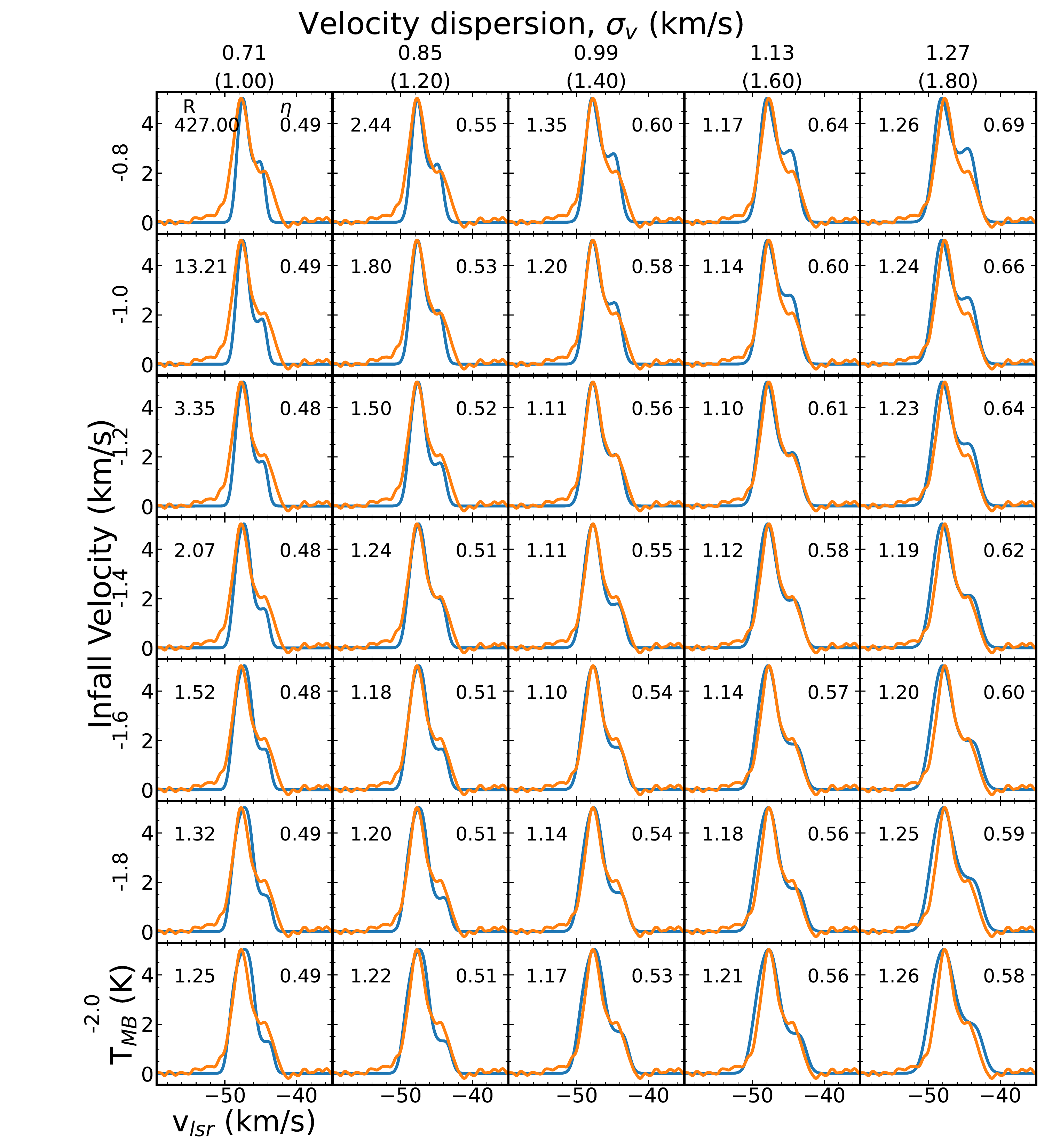}
\caption{Comparison of a grid of LIME models with the emission at the central pixel. The models have a uniform infall velocity ($f=1$, so \vin$=v_{\rm c}$) in the clump with a $p=2$ density profile and a uniform gas temperature of 20\,K. In each sub-panel, the orange curve shows the observed spectrum at the central pixel and the blue curve shows the model of the indicated velocity dispersion and \vin. The value in the upper right of each sub-panel shows the scaling factor applied to the models ($\eta$) to match the peak 
of the observed line. The other value in each sub-panel shows the $R$ statistic for the fit. For ease of display, the figure only shows a sub-sampled grid of the models, stepping by 0.2\,\kms\ in both infall velocity and \dopb\ whereas the complete grid of models is sampled at 0.1\,\kms\ in both parameters. }
\label{fig: LIME_grid}
% Notebook: ViewGrid.ipynb 
% file like: /Users/mccssgf/RadiativeTransfer/LimeInfallModels/Grid/new_model-subgrid_obs_Grid5e_dw_ref-14_14_sm.pdf
\end{figure}

\begin{figure} 
   \centering
   \includegraphics[width=\columnwidth]{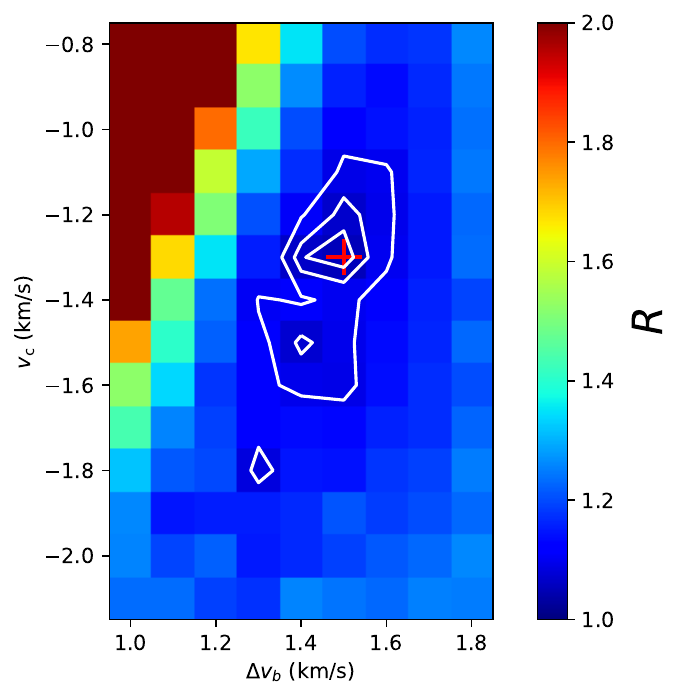}
\caption{The $v_{\rm c}-$\dopb\ plane for the grid of $f=1$ models shown in Figure~\ref{fig: LIME_grid}. The axes show the full range of 
\vin($=v_c$ for these $f=1$ models) and \dopb\ explored. Both values were sampled every 0.1\,\kms. The red cross marks  the best fit model ($M_{LIME,7}$) which has   \vin=-1.3\,\kms, \dopb=1.50\,\kms\  (\disp= 1.06\,\kms) and $R=1.05$. The lowest contour shown has $R=1.10$.}
% Notebook: ViewGrid.ipynb 
\label{fig:vin_db-plane}
\end{figure}

To find the best fit infall velocity $v_{\rm in}$ and velocity dispersion, $\sigma_v$, we ran grids of models with different infall velocities and velocity dispersions. %Initially we explored models with a constant infall velocity across the clump and a constant temperature, which are similar to those in \citet{2013A&A...555A.112P}. 
LIME describes the velocity width of the gas in terms of the Doppler-$b$ parameter, \dopb$=\sqrt{2}$\disp, where \disp~is the velocity dispersion. The grids of LIME models were sampled at 0.1\,\kms\ in \vin\ and 0.1\,\kms\ in \dopb. An example grid of LIME models is shown in Figure~\ref{fig: LIME_grid}, corresponding to model $M_{\rm LIME,7}$ in Table~\ref{tab:bestFitModels} ($p=2.0$, $f=1$, and a constant temperature $T=20\,\mathrm{K}$). Figure~\ref{fig:vin_db-plane} presents contours of the $R$ parameter on the $v_{\rm c}$--$\sigma_v$ plane.\footnote{The $v_{\rm c}$--$\sigma_v$ planes for the other $p=2$ LIME models are shown in Figure~\ref{fig:other_vin_db-plane}.  For this model, we find that the best fit for the central pixel is obtained with $v_{\rm in}=-1.3\,\mathrm{km\,s^{-1}}$ and $\sigma_v=1.06\,\mathrm{km\,s^{-1}}$ ($\Delta v_b=1.50\,\mathrm{km\,s^{-1}}$). Similar $v_{\rm c}$--$\sigma_v$ planes of $p=2$ models are shown in Figure~\ref{fig:other_vin_db-plane}.}

% 1 Aug2022
%Best fit models uniform vin, T=20K:
% p=1.5 - two positions in grid: vin =-1.3, dop = 1.5 and vin = -1.4, dop = 1.4

% \begin{figure} 
%    \centering
%    \includegraphics[width=\columnwidth]{new_model-zoomgrid_obs_Grid6e-p_2-20_dw_ref-13_13_sm.pdf}
% \caption{Grid of models around the best fit model for a uniform infall velocity model with a gas temperature of 20\,K. {same as Figure 9?}This is the same grid of models shown in Figure~\ref{fig: LIME_grid}.}
% \label{fig:zoom-LIME_grid}
% % Notebook: ViewGrid.ipynb 
%\end{figure}

For direct comparison with \citet{2013A&A...555A.112P}, we also explored models with $p=1.5$ and a uniform temperature of 20\,K (see $M_{LIME,2}$ in Table~\ref{tab:bestFitModels}). With these parameters, the best fit has an infall velocity of $-1.3$\,\kms and Doppler width of 1.5\,\kms (corresponding to a dispersion of 1.1\,\kms), the same as for the $p=2$ model ($M_{LIME,7}$). 
\citet{2013A&A...555A.112P}
found a similar velocity dispersion of 1.0\,{\kms}, but a smaller in magnitude infall velocity of $0.7$\,\kms. However, it should be noted that they were also fitting a different dataset taken with the Mopra telescope. In addition, \citet{2013A&A...555A.112P} used RATRAN, rather than  LIME (Sec.~\ref{subsub:cmc}).

Comparing uniform infall velocity ($f=1$) models with a temperature profile ($M_{LIME,1}$). with those with 20\,K throughout (all with $p=1.5$) ($M_{LIME,2}$), shows that the presence of a temperature gradient results in an infall velocity 0.2\,\kms\ smaller in magnitude and a velocity dispersion 0.1 \,\kms\ larger (Table~\ref{tab:bestFitModels}). However, we do not consider these significant differences given the uncertainties in the fitting process and the underlying assumptions, and so our results suggest that density profiles have a negligible impact on the profiles.

A single model with twice the outer radius of the other models was also run to test for any influence of the size on the inferred infall properties. The best fit values from this model ($M_{LIME,8}$) are very similar to the models with the smaller maximum radius, although the line peak temperature is higher. 

%\label{tab:bestfit_central_pixel}
%\input{table_bestfit_central_pixel.tex}

%\begin{figure} 
 %  \centering
  % \includegraphics[width=\columnwidth]{T20_central_pixel_profile.pdf}
%\caption{The central pixel line profiles from different models compared to the observational data. The light gray lines represent the HCO$^{+}$ $J\,=$1--0 data at the central pixel. The red lines are the central pixel profiles from models. The dashed black lines represent the differences between the models and observational data. The dotted horizontal lines represent where the differences are zero. The profiles of RT modelling (RATRAN, LIME, and RADMC-3D) are fitted to the telescope main beam efficiency and normalised to fit to the data. }
%\label{fig: bestfit_central_pixel}
%\end{figure}

%To investigate the sensitivity and accuracy of Hill5 model, we fitted the best fit spectra at the central pixel from the RT models with Hill5. The resulting infall parameters are listed in Table~\ref{tab:Hill_models} which should be compared with the input parameters listed in Table~\ref{tab:bestfit_central_pixel}. Although the Hill5 model fit the model line profiles (Figure~\ref{fig: Hill_models}), it always fits with smaller infall velocities than the input values in the models. A more detailed discussion is given in sect~\ref{subsec:Hill_accuracy}. 

\begin{figure} 
   \centering
   \includegraphics[width=\columnwidth]{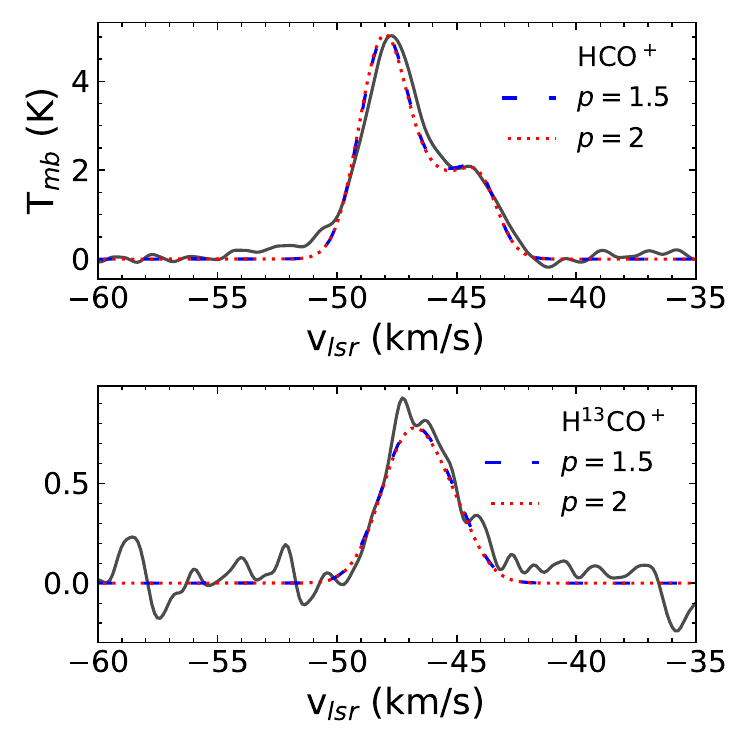}
\caption{The central pixel profiles from two different density profiles in LIME models compared to the observational data. The models used temperature profiles with a minimum temperature of 20\,K. 
{\it Left:} \HCOp\,J$=1-0$ and {\it Right:} \HTCOp\,J$=1-0$. In each panel, the grey continuous line shows the observed emission at the central pixel. The red and blue broken curves show the best fit models for the density profiles with $p=1.5$ and $p=2$ with uniform infall velocities ($f=1$). For both models, the infall velocity is $-1.1$~\kms\ and \dopb =1.6\,\kms. 
% GAF ToDo still - 6 Oct 2024
%{generate H13CO+ with different velocity profiles. }
}
\label{fig:lime-models-central_pixel}
% Notebook: FinalPlots.ipynb
\end{figure}

Figure~\ref{fig:lime-models-central_pixel} compares the best fit models from grids of models with $p=1.5$ ($M_{LIME,1}$) and $p=2$ ($M_{LIME,6}$) density profiles (and matching temperature gradients). 
The upper panel of Figure~\ref{fig:lime-models-central_pixel} shows the \HCOp\ while the lower panel shows \HTCOp. The two different density profile models have the same best fit values for the infall velocity and velocity dispersion, however, the lines they produce have different line peak intensities. The peak intensity of the \HCOp\ line from the  $p=1.5$ model ($M_{LIME,1}$) is 7.5\,K while it is 8.7\,K for the $p=2$ model ($M_{LIME,6}$). This is not apparent from the figure as the models are plotted adopting an efficiency factor to scale the model to the observed line peak intensity. Similarly for the \HTCOp\ where the $p=1.5$ model ($M_{LIME,1}$) predicts a weaker line peak temperature (1.2\,K), than the $p=2$ model ($M_{LIME,6}$) with a peak temperature of 1.9\,K.

%\begin{figure} 
 %  \centering
  % \includegraphics[width=\columnwidth]{GAF_all_models_central_pixel.pdf}
%\caption{Comparison of the model central pixel spectra from RATRAN, LIME, and RADMC-3D for the best fit values in Table \ref{tab:bestFitModels} for a temperature of 20K. The models have all been normalised to the peak intensity of the observed line. }
%\label{fig: ratran_lime_radmc3d_central_pixel}
%\end{figure}

\begin{figure} 
   \centering
   \includegraphics[width=\columnwidth]{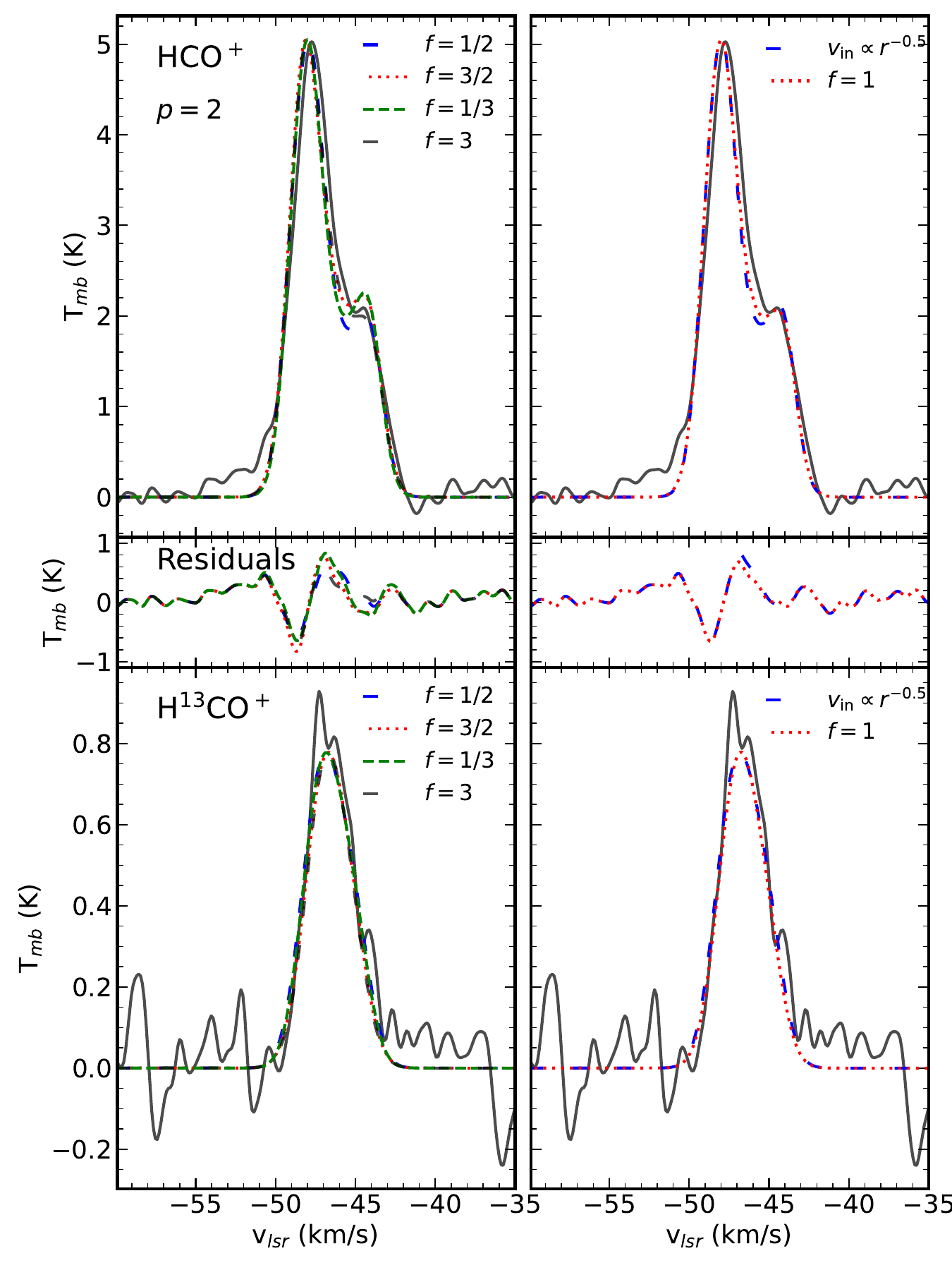}
\caption{The central pixel \HCOp\,J$=1-0$ and \HTCOp\,J$=1-0$ profiles from different best fit models with a $p=2$ density profile compared to the observational data. The top row shows a comparison of six models with different infall velocity profiles. The parameters of these models are given in Table~\ref{tab:bestFitModels}. The middle row shows the difference between each model and observations. The bottom row shows the predicted \HTCOp\ from these same models compared with the observed \HTCOp\ emission. 
% GAF ToDo - 6 Oct 2024
%{correct the right panel label p=1. }% what actually is vff model for H13CO+.
}
\label{fig:lime-models-vprof-central_pixel}
% Notebook: FinalPlots.ipynb
\end{figure}

\begin{figure}
    \centering
    \includegraphics[width=0.5\textwidth]{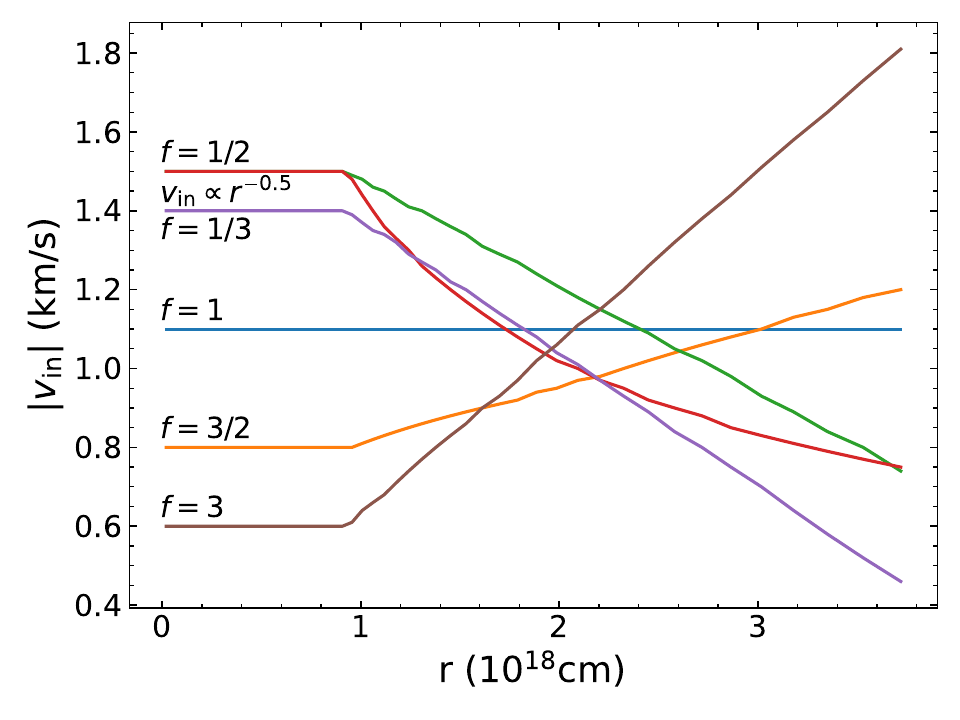}
    \caption{Velocity profiles of the models with $p=2$ which give the best fit to the central pixel.
%    {Make consistent with table 1 and Fig 16. Add labels for the curves.}
    }
    \label{fig:vprofiles}
    % Plot from BFProfilesPlot.ipynb
\end{figure}

Figure~\ref{fig:lime-models-vprof-central_pixel} shows the observations compared with the best fit models for five models with a range of values for $f$ from $1/3$ to $3$ while $p=2$, i.e., $M_{LIME,4}$, $M_{LIME,5}$, $M_{LIME,6}$, $M_{LIME,9}$, and $M_{LIME,10}$. Figures~\ref{fig:lime-models-central_pixel} and \ref{fig:lime-models-vprof-central_pixel} show that all these models can reproduce the emission at the central pixel both for the \HCOp\ and \HTCOp\ to a similar level, with the best fit models having $R$ values between 1.08 (for the $M_{LIME,3}$ model) and 1.13 (for several models). Figure~\ref{fig:vprofiles} shows the adopted radial infall velocity profiles, where the parameter $f$ controls whether the infall velocity decreases ($f<1$), remains constant ($f=1$), or increases ($f>1$) with radius while matching the same velocity at the central-pixel radius. As Figures~\ref{fig:lime-models-central_pixel}, \ref{fig:vprofiles}, and Table~\ref{tab:bestFitModels} show the choice of the structure of the infall velocity has a significant impact on the derived best fit \vin\ at the radius of the central pixel, $v_c$. Models with larger infall velocities in the outer regions of the clump require smaller infall velocities in the central region. Looking at the $p=2$ models in Table~\ref{tab:bestFitModels}, the inferred $v_c$ increases in magnitude from 0.8\,\kms\ for $f=3$ ($M_{LIME,10}$) to 1.4\,\kms\ for $f=1/3$ ($M_{LIME,4}$). Fitting with a uniform infall velocity throughout the clump ($f=1$, $M_{LIME,6}$) implies a central $v_c$ of $-1.1$\,\kms\ intermediate between the case where the infall velocity increases with radius and that where it decreases.

\subsubsection{Comparison of Modelling Codes}\label{subsub:cmc}

\begin{figure} 
   \centering
   \includegraphics[width=\columnwidth]{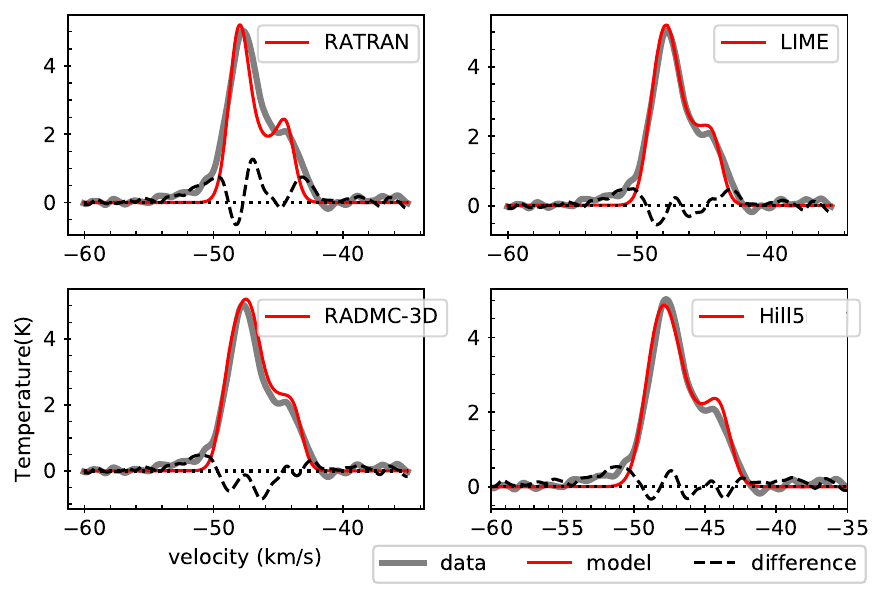}
\caption{The central pixel line profiles from different models compared to the observational data. The light gray lines represent the HCO$^{+}$\,J$=1-0$ data at the central pixel. The red lines are the central pixel profiles from models. The dashed black lines represent the differences between the models and observational data. The dotted horizontal lines represent where the differences are zero. The profiles of RT modelling (RATRAN, LIME, and RADMC-3D) are fitted to the telescope main beam efficiency and normalised to fit to the data. }
\label{fig: bestfit_central_pixel}
\end{figure}

To investigate the impact of using different RT to model the infall, in addition to LIME we used RATRAN and RADMC-3D to find the best fit parameters for models with uniform infall velocity, $p=1.5$ and uniform temperatures of 20\,K (using a grid-search method). The results for these models are also shown in Table~\ref{tab:bestFitModels} and these can be directly compared to the $p=1.5$, $f=1$ (20\,K) models in the table. 
A comparison of the best fit results are presented in Figure~\ref{fig: bestfit_central_pixel}. For comparison, the result of applying Hill5 to the central pixel is also shown in  Figure~\ref{fig: bestfit_central_pixel} and Table~\ref{tab:bestFitHill5Models}. 

\begin{figure} 
   \centering
   \includegraphics[width=\columnwidth]{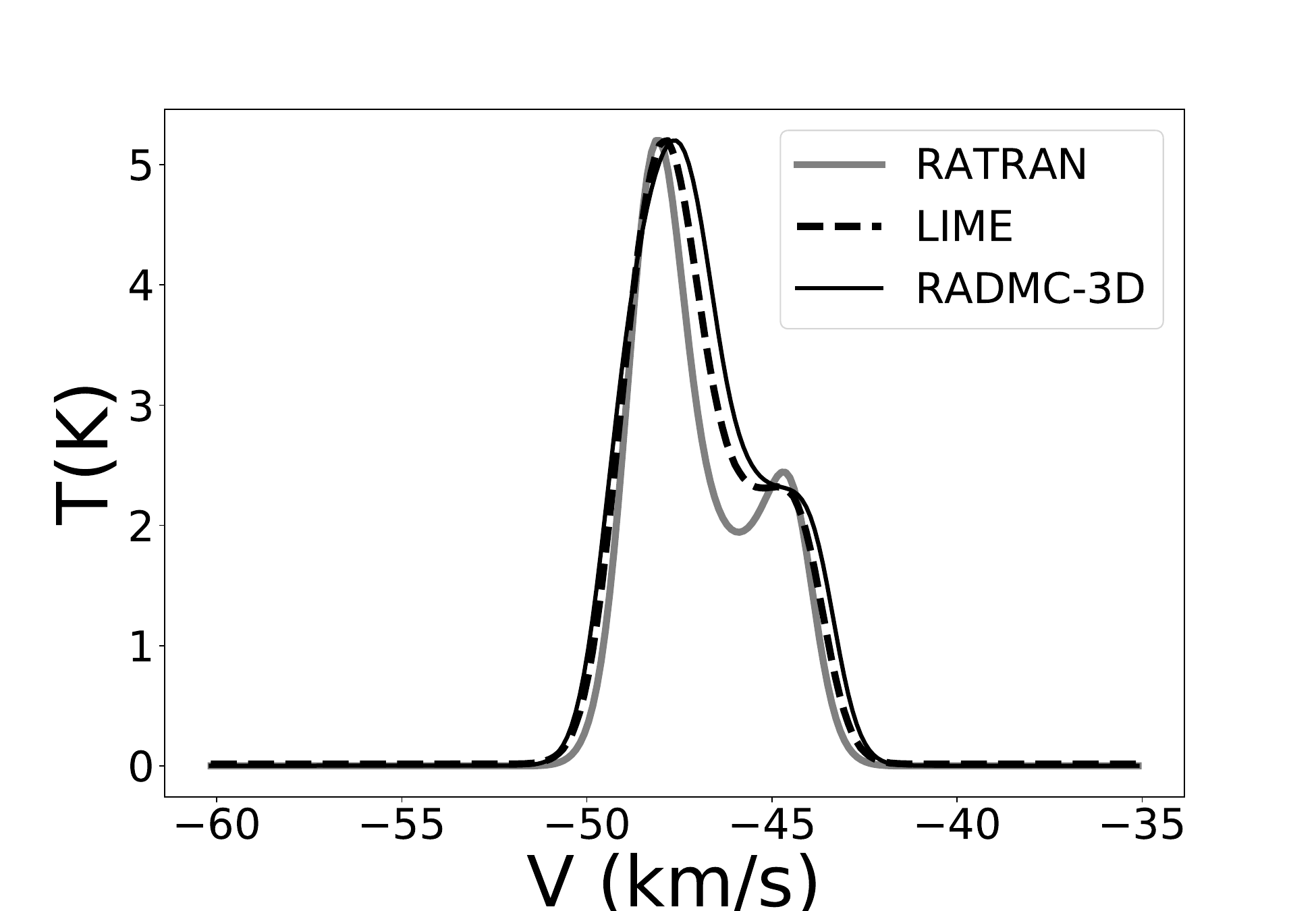}
\caption{Comparison of the model central pixel spectra from RATRAN, LIME, and RADMC-3D for the best fit values in Table \ref{tab:bestFitModels} for a temperature of 20\,K. The models, i.e., 
%$M_{LIME,2}$, 
$M_{LIME,7}$, $M_{RADMC3D}$, and $M_{RATRAN}$ have all been normalised to the peak intensity of the observed line. }
\label{fig: ratran_lime_radmc3d_central_pixel}
\end{figure}

The best fit results from the models are surprisingly different though all of them can fit the data reasonably well and to similar degrees of accuracy. The infall velocities from the three RT models, RATRAN, LIME, and RADMC-3D, span a range of a factor of 2, from --0.8\,\kms\ (RATRAN) to --1.6\,\kms (RADMC-3D). The spread in the inferred velocity dispersion, 0.3\,\kms, is smaller but still significant while there is also a more than 1\,K difference in the predicted line peak intensity. In each case, RADMC-3D produces the highest line intensity which may be related to the fact that the radiative transfer for the line emission in RADMC-3D is evaluated in the large velocity gradient approximation which results in a lower optical
depth at a given velocity and so requires a large velocity to match the observed line width while also tracing higher excitation temperature regions of the model. The comparison of the line shape at the central pixel of these models 
%($M_{LIME,2}$, 
($M_{LIME,7}$, $M_{RADMC3D}$, and $M_{RATRAN}$) is shown in Figure~\ref{fig: ratran_lime_radmc3d_central_pixel} which shows that RATRAN produces a line profile with a larger contrast between the temperature of the blue peak and the minimum between the red and blue peaks (so a larger value of $D_r$, Appendix~\ref{sec:linechara}) than either LIME or RADMC-3D. 
The infall velocity from LIME and inferred by Hill5 map are in good agreement, although Hill5 implies a smaller velocity dispersion. 
%{check and update after final ratran grid run for Table 1 and RATRAN results need explanation.}

\subsection{The Spatial Distribution of Line Profiles}\label{sec:spatial}
 
\begin{figure*} 
   \centering\includegraphics[width=0.65\textwidth,height=0.60\textwidth,trim={{0.1\textwidth} {0.2\textwidth} {0.1\textwidth} {0.1\textwidth}},clip]{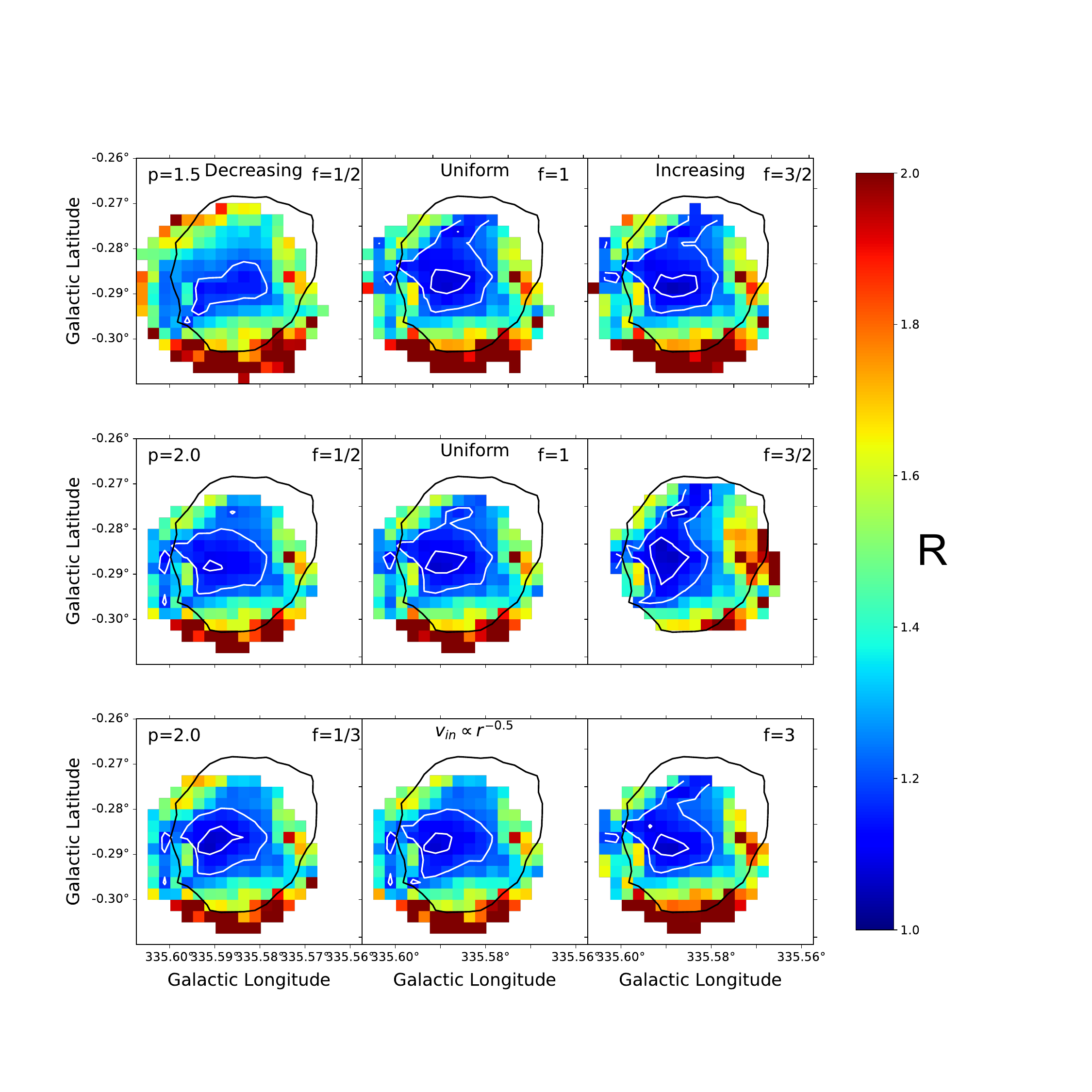}
\caption{The spatial distribution of the $R$ statistic in the best fit models and observations. The top row shows models with $p=1.5$ and the other two rows models with $p=2.0$. The columns show models with $f$ from a decreasing \vin\ profile ($f<1$, in the left columns), a uniform \vin\ (the middle columns), and an increasing \vin\ profile (f>1, the right columns). The white contours show $R=1.1$ and $1.2$, and the single black contour shows where the integrated intensity of the emission over the velocity range used to calculate $P_r$ is 20\% of its peak intensity.}
%The number of pixels which are well fitted by each model is listed in Table~\ref{tab:spatial_pixel_num}. 
%{change labels for central and top panels. f=1.}}
\label{fig:spatial_distribution}
% Figure from PlotSpatialDistFit.ipynb
\end{figure*}

For the LIME models which fit the central pixel spectra best among their own series of parameters, we also compare the spatial distribution of the models with the data. 
The spatial distribution of $R$ statistic in those models is summarised in Figure~\ref{fig:spatial_distribution}. For this comparison, the effective telescope efficiency is calculated for each pixel which allows for possible unresolved changes in the structure of the clump. %Table~\ref{tab:spatial_pixel_num} summarises these images by listing the statistics for the spatial pixels which are well fitted by each model. 

Overall, in the central regions, all the models appear to provide reasonable fits to the observations, although models that have infall velocities that decrease with radius ($f<1$) appear to produce slightly worse agreement (fitting a smaller number of pixels with small $R$ values) than models with uniform or increasing ($f\ge1$) infall velocity.
However, the models that have infall velocities that increase with radius ($f>1$) provide fits with lower $R$ values to the emission to the north of the map. 
%The models here have a fixed central velocity and so changes in the central velocity of the emission across the region could contribute to a mismatch between the model and observations. 

%\begin{figure} 
%   \centering
%   \includegraphics[width=\columnwidth]{nico_T20_spatial_distribution.pdf}
%\caption{The spatial distribution fitting from RATRAN and RADMC-3D. The gray lines represent the observational data, while the red profiles represent the line profiles from modelling. The upper panel is from RADMC-3D modelling. The lower panel is from RATRAN modelling. The central pixel is labelled with the black dot. RADMC-3D tends to fit the observational data in all directions than RATRAN from appearances. The intensities of both models are converted to match the telescope main beam efficiency. }
%\label{fig: spatial_distribution}
%\end{figure}

\begin{figure} 
   \centering
   \includegraphics[width=\columnwidth]{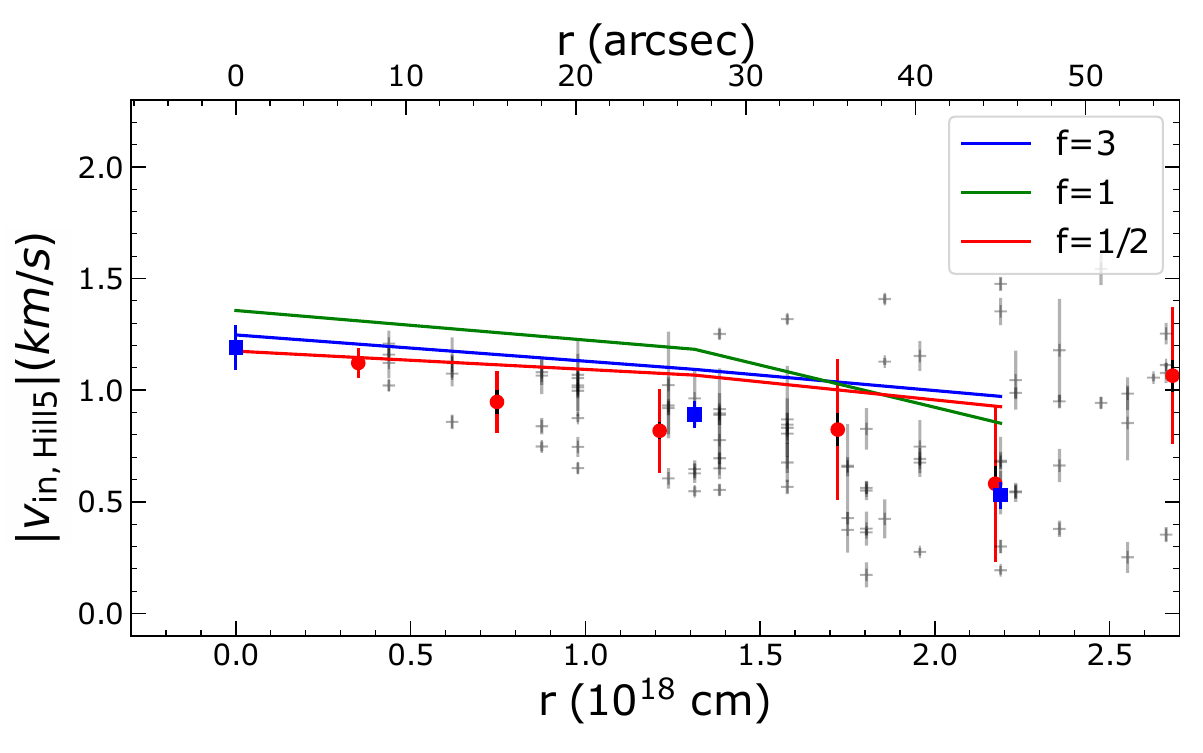}
\caption{Infall velocity \vin\ (v$_{\rm in,Hill5}$) from  Hill5 fitted spectra as a function of radius from the central pixel. 
 The grey and red points are as shown in Figure~\ref{fig:hill5Rad}.
 The three blue points show the result of Hill5 fits to the observations at the central pixel and the observations averaged over the two annular regions A1 and A2.
The three colour lines connect the Hill5 fits to the annular average spectra at three radii (the central pixel and the spectra averaged over the two annular regions A1 and A2) from the best fit models for $f=1/2, 1\ \mathrm{and}\ 3$ (models $M_{LIME,5}$, $M_{LIME,6}$ and $M_{LIME,10}$ respectively). }
\label{fig:hill5RadMod}
% From notebook: PaperFigures/Hill5vin-dist-plot
\end{figure}

Figure~\ref{fig:hill5RadMod} shows the region of Figure~\ref{fig:hill5Rad} where $r<2.7\times10^{18}$\,cm with 
the results of Hill5 fits for \vin\ (v$_{\rm in,Hill5}$) to individual pixels in the \HCOp cube together with the binned average of these values. In addition, the values  of \vin\ determined by Hill5 fits to spectrum at the central pixel, and averaged over the two annular regions A1 and A2 (Table~\ref{tab:annuli}) are shown.
All three of these estimates of \vin\ are in good agreement, showing a decrease in the magnitude of \vin\ from about 1.2 \kms\ at the centre of the map to $\sim0.5$\,\kms\ at a radius of $2.2\times10^{18}$\,cm (45''). The solid curves on the figure show the result of Hill5 fitting to the central pixel and the annular regions for the $p=2.0$, $f=1/3, 1$ and $3$ models ($M_{LIME,4}$, $M_{LIME,6}$, and $M_{LIME,10}$) which provide the best fit to the central pixel. 
Although only one of these models has an infall velocity that actually decreases with radius (the model with $f=1/3$ ($M_{LIME,4}$)), the Hill5 fits for all three models show a decrease in the fitted infall velocity with radius, all approximately consistent with the observations. For the same values of $f$, other models with slightly different velocities and dispersions (and so slightly worse fits to the central pixel) can come closer to the observed values and trend.

The apparent consistency of these different models shows that  the apparent decrease in \vin\
inferred by Hill5 for the observations does not necessarily reflect an actual decrease in the infall velocity. Rather this decrease is likely a result of a combination of decreasing optical depth and geometry leading to a less asymmetric line profile and resulting in Hill5 inferring a lower \vin.

\subsection{\HTCOp}
\label{sec:h13cop_dv}

Observations of \HTCOp\,J$=1-0$ can also provide information about the spatial variation in the infall velocity profile. Its lower optical depth, due to its lower abundance than the main isotopologue, results in this species not having a significant blue asymmetric line profile. On the other hand, systematic motions such as due to infall, should be reflected in
the width of the line. However,  due to weakness of the line and so the low signal to noise ratio in the Mopra observations,  analysis of the existing data is limited to 
the study of the radial profile of the \HTCOp\ line width, rather than its full spatial distribution over the region.

\begin{figure*}
\centering
\includegraphics[width=\columnwidth]{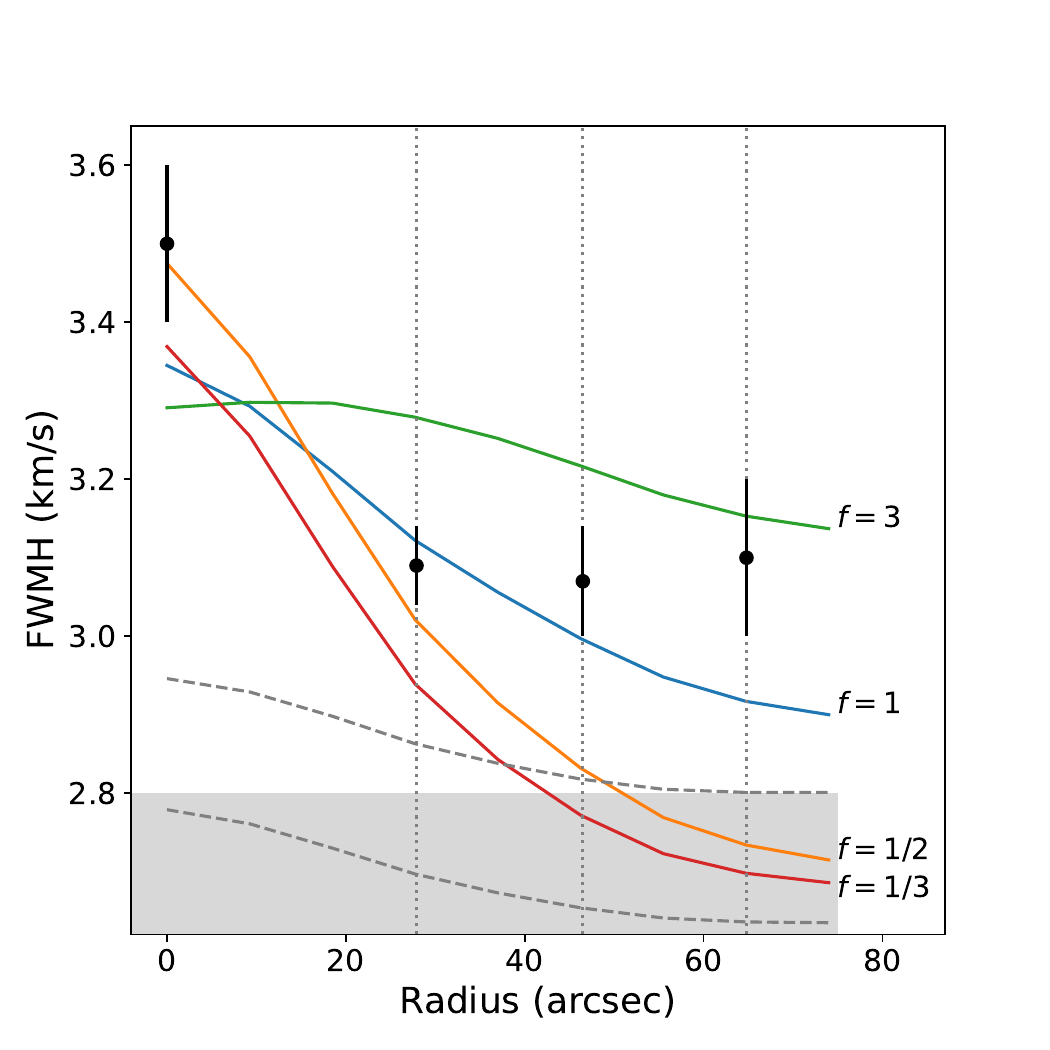}\includegraphics[width=\columnwidth]{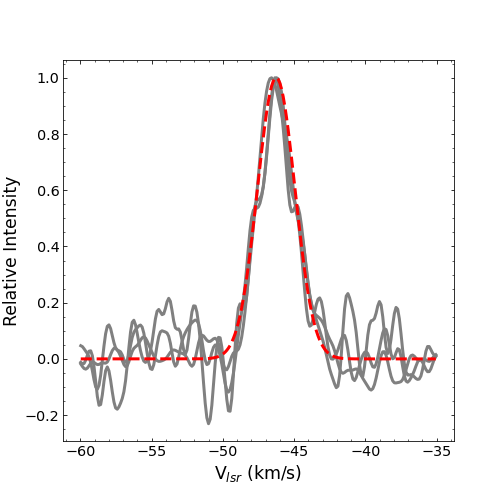}
\caption{{\it Left:} The variation of the width of the H$^{13}$CO$^+$\,J$=1-0$ line for models with different \vin\ radial structures. The $p=2$ best fit models are shown, labelled by the $f$ value for the model. The vertical dotted lines show the radii of the annular apertures used to average the observed spectra shown in the accompanying plot. The black points and error bars show the observed FWHM and their uncertainties determined from Gaussian fits to the H$^{13}$CO$^+$\,J$=1-0$ line at the centre of the map and the annular average spectra. The gray-shaded region shows the range of the intrinsic velocity dispersion (expressed as FWHM) for the infall models shown. The dashed gray curves show the FWHM of the line from two static models (models with no infall), demonstrating the expected variation in the absence of infall. The velocity dispersion in these models was selected to span the range of values in the models shown by the coloured curves. {\it Right:} The observed H$^{13}$CO$^+$\,J$=1-0$ averaged over the annular regions with mid-point radius shown on the panel to the left (grey). The red dashed curve shows the Gaussian fit to the spectrum at radius 65''. 
%{f=0 to f=1.} 
} \label{fig:thin-dv}
% Notebook: Figure from GaussianFitCube.ipynb
% H13CO+ data fits from AnnAverSpec
\end{figure*}

 Figure~\ref{fig:thin-dv} (left) shows the FWHM of \HTCOp J$=1-0$ derived from Gaussian fits to the observed spectra at the central pixel and in the three annular regions around the central pixel (Table~\ref{tab:annuli}). 
The FWHM peaks towards the central pixel while at larger radii it is smaller, but remarkably uniform at $\sim3.1$ \kms\ even out to 65'' from the central source. This is clearly shown in the right-hand panel of the figure which shows the normalised \HTCOp\ spectra in the annular regions.  
Comparing the observed \HTCOp\ widths with the intrinsic velocity width of the gas in the models (as shown by the gray-shaded region on Figure~\ref{fig:thin-dv} (left) and the width expected for models with no infall (the gray-dashed curves), 
the figure shows that at all positions the \HTCOp\ has a line width greater than the width necessary to fit the \HCOp\ emission. This points to the infall motions in the clump making a significant contribution to the width of the \HTCOp\ emission.

The variation of the predicted \HTCOp\ line width with radius from the models with infall is shown by the coloured curves demonstrating the impact of the infall on this line. All the infall models show a decrease in line width with increasing radius, but the models in which the infall velocity decreases with radius ($f=1/2$ and $f=1/3$) show the steepest decrease with radius. At the outer edge of these models, the line width falls to close to the intrinsic velocity width in the models. Although these models match the central line width and the decrease at 28'', they can not reproduce the uniform line width beyond 28''. 

Models where the infall velocity is uniform, or increases with radius ($f=1$ and $3$) show a slower decrease in the width of \HTCOp\ with increasing radius, with the $f=3$ model showing a variation of only a little more than 0.1 \kms\ over the full range of radius. Although this model does not match the large width at the centre as closely as $f=1$ and $1/2$ models do, the $f=3$ model comes closest to matching the near uniform observed line width away from the centre.

In summary, infall appears to make a significant contribution to the observed line widths of \HTCOp\,J$=1-0$. Models with a decrease in \vin\ out to $\sim30''$ match the drop in line width close to the central source, however models where the infall velocity is uniform or increasing with radius best match the observed spatial distribution of the \HTCOp\ line width at the largest radii.

\section{Discussion} \label{sec:discussion}

\subsection{The Impact of The Density Profile} 

Our models show that the assumed density profile ($p=1.5$ or $p=2.0$) has a relatively small impact on the best fit infall velocity for a given variation of the infall velocity with radius. For the uniform infall velocity case ($f=1$), for both models with a constant temperature (20\,K, $M_{LIME,2}$ and $M_{LIME,7}$) and those with a temperature profile ($M_{LIME,1}$ and $M_{LIME,6}$), the best fit values for \vin\ are the same for both density profiles. For a uniform temperature of 20\,K ($M_{LIME,2}$ and $M_{LIME,7}$), the infall velocity is $-1.3$\,\kms and \dopb = 1.5\,\kms\ while models with a temperature profile ($M_{LIME,1}$ and $M_{LIME,6}$) have a somewhat smaller infall velocity ($-1.1$\,\kms) and slightly larger \dopb = 1.6\,\kms. There is a similar difference of 0.2\,\kms\ in \vin\ for the two density profiles for models with $f=1/2$ and $f=3/2$ ($M_{LIME,0}$ and $M_{LIME,5}$ for $f=1/2$, and $M_{LIME,3}$ and $M_{LIME,9}$ for $f=3/2$ in Table~\ref{tab:bestFitModels}). The choice of density profile does, however, impact the predicted line strength. For example, comparing the $f=1$ models $M_{LIME,1}$ and $M_{LIME,6}$, the model with $p=2$, $M_{LIME,6}$, has a central line peak intensity of 8.7\,K compared with 7.5\,K for the model with $p=1.5$, $M_{LIME,1}$.

% Spatial fit statistics - currently not in the paper
%\input{table_spatial_stats.tex}

\subsection{Infall Velocity Profile} 

\begin{figure*} 
   \centering
   \includegraphics[width=\textwidth,clip]{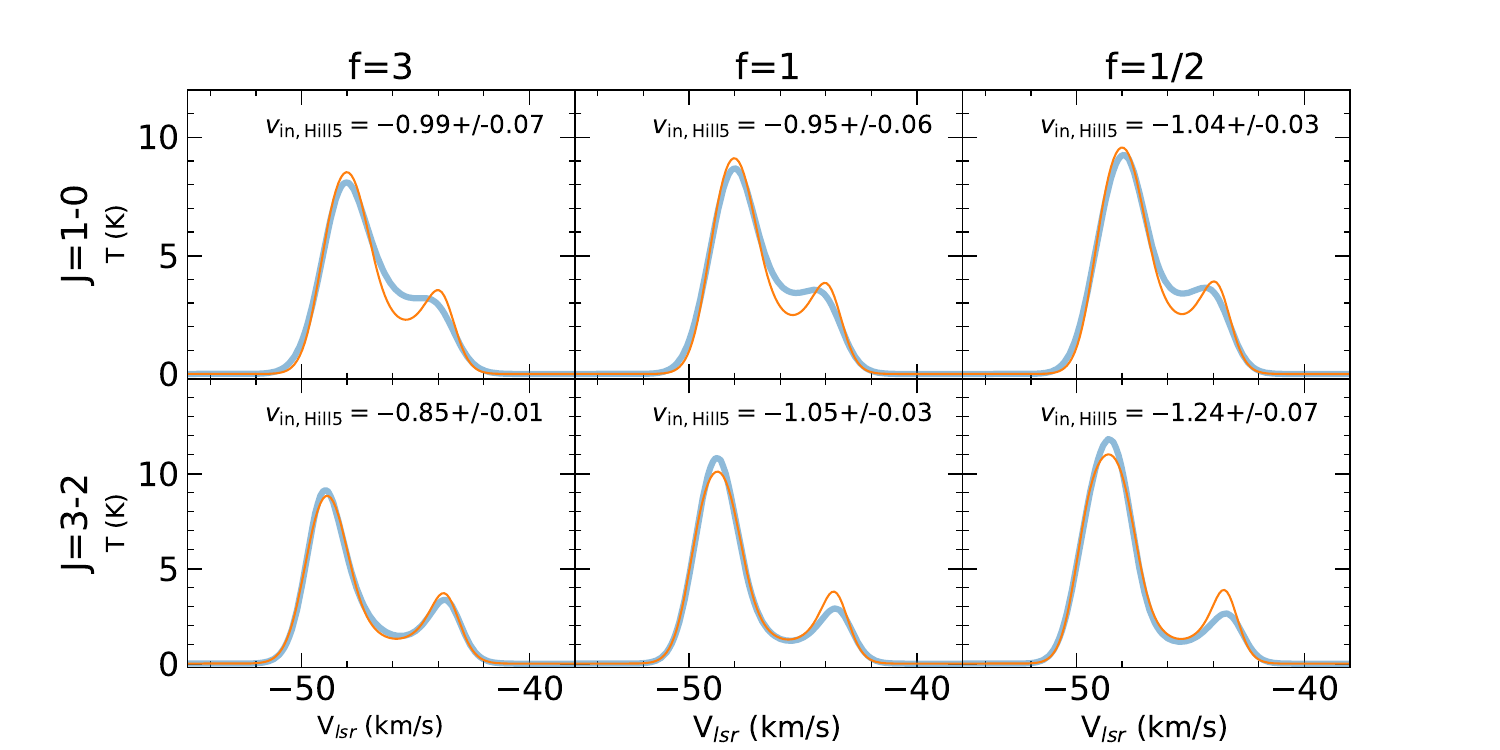}
\caption{Comparison of the \HCOp\,J$=1-0$ (upper) and J$=3-2$ (lower) spectra towards the central pixel of the LIME $f=3$ (left), $f=1$ (middle) and $f=1/2$ (right) models ($M_{LIME,10}$,$M_{LIME,6}$, and $M_{LIME,5}$, respectively). The broad, blue curve in each panel shows the model spectrum and the thinner orange curve, the Hill5 fit to the spectrum. The Hill5 fitted infall velocity and its uncertainty are shown in the upper right of each panel. Note that for these fits no noise was added to the model spectra. The model J$=3-2$ spectra have been convolved with a 22'' FWHM beam and while the J$=1-0$ transitions have been convolved with the 37'' beam of the observations.} \label{fig:higherJcomp}
%
% Figure from Python/PaperFigures/Compare_Transitions.ipynb
% Plots are written to Python/Compare_Transitions_Plot
\end{figure*}

Applying Hill5 to the \HCOp\,J$=1-0$ cubes produces a map of the inferred \vin\ which suggests that close to the source, the infall velocity decreases with increasing distance from the source. However, comparison with radiative transfer models shows 
 this apparent decrease can be consistent with a range of radially varying infall velocity, from decreasing to increasing with radius (Sec.~\ref{sec:spatial}). The degeneracy between these models could however be broken with observations of additional transitions of \HCOp. 
 As an example, Figure~\ref{fig:higherJcomp} shows a comparison of the J=1--0 transition towards the centre of the three models shown in Figure~\ref{fig:hill5RadMod}.
 Compared with the J$=1-0$ transitions, the J$=3-2$ transitions show a stronger blue asymmetry. This is the result of the difference in the excitation of the transitions plus the smaller telescope beam used to convolve the models (22'' for the J$=3-2$ transition, representative of the beam of the 12-m diameter APEX telescope compared with the 37'' for the J$=1-0$ transition.). 
 The figure shows that the infall velocity estimated by Hill5 from the J$=1-0$ transitions differs at most by 0.12\,\kms\ between the models. On the other hand, higher angular resolution observations of the J$=3-2$ transition seem to provide a better discrimination between the models, with the magnitude of the infall velocities ranging from 0.86\,\kms\ to 1.25\,\kms, a range of 0.39\,\kms.

As discussed in Section~\ref{sec:h13cop_dv}, infall motions affect the  line width of the optically thinner \HTCOp\,J$=1-0$. 
If the line width at the centre of the clump is not significantly
impacted by other motions, such as the impact of the outflows from
the embedded sources, the decrease in line width from  3.5~\kms, at the centre of the clump to $\sim3.1$~\kms\ at larger radii seen in Figure~\ref{fig:thin-dv} would appear most consistent with a model where the infall velocity is greatest closer to the embedded sources. However, the uniform line width at larger distances suggests higher infall velocities at these larger distances. Note that, except at the central position, these line widths are measured in spectra averaged over annular regions and so sample relatively large volumes of the clump, meaning that they are unlikely to have significant contributions from outflows as the outflows are of limited spatial extent \citep{2021A&A...645A.142A}. Also, the clump is seen in absorption at 8\,$\mu$m \citep{2013A&A...555A.112P}, indicating that it has not been strongly disrupted by outflows, which would otherwise reduce the extinction and allow near-infrared emission to escape.

\subsection{Hill5 Model Sensitivity and Accuracy}     \label{subsec:Hill_accuracy} 
%{GAF: Could still do with a bit of work.}

\begin{table}
\centering
\caption{Results from applying Hill5 to the central spectra from the 20 K RT models ($M_{LIME,2}$ and $M_{LIME,7}$) shown in Table~\ref{tab:bestFitModels}.}
\label{tab:Hill_models}
\resizebox{0.45\textwidth}{!}{%
\begin{tabular}{lccc}
\hline
Infall parameters & RATRAN & LIME & RADMC-3D \\
\hline
$\sigma$ ({\kms}) 
& {1.06$\pm$0.01 (1.13)} 
& {1.02$\pm$0.02 (1.06)} 
& {1.02$\pm$0.01 (1.30)} \\
v$_{\mathrm{in,Hill5}}$ ({\kms})
& -0.80$\pm$0.02 (0.9)
& -1.16$\pm$0.04 (1.30)
& -1.21$\pm$0.03 (1.60) \\
\hline
\multicolumn{4}{p{1.15\columnwidth}}{Notes. The values in brackets indicate the actual values from the models in Table~\ref{tab:bestFitModels}.}
\end{tabular}}
\end{table}

The Hill5 method was developed for the emission from regions where $|$\vin$|<$\disp\  but it can give good estimates of infall velocity when \disp\ and \vin\ are comparable \citep{2005ApJ...620..800D, 2010MNRAS.402...73B}. However, 
it has been suggested that the `Hill' method underestimates the actual infall velocities in a region in some circumstances \citep[e.g.][]{2006ApJ...637..860W}. A possible reason could be that the `Hill' model, as well as its precursor `two-layer' model, do not take into account the portion of the cloud whose systematic line-of-sight velocity is zero \citep{2006ApJ...637..860W}. 

\begin{figure} 
   \centering
   \includegraphics[width=0.30\columnwidth]{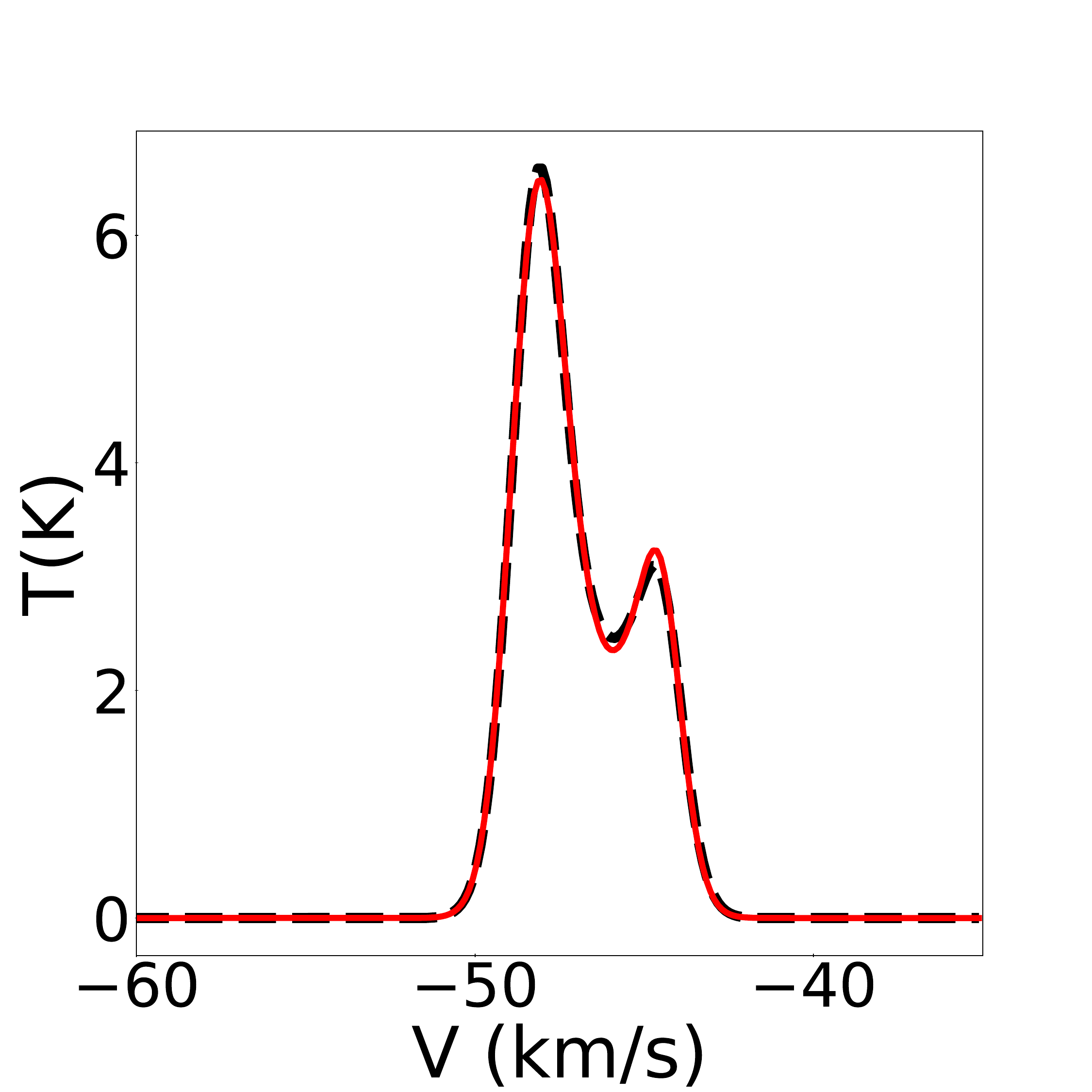}\includegraphics[width=0.30\columnwidth]{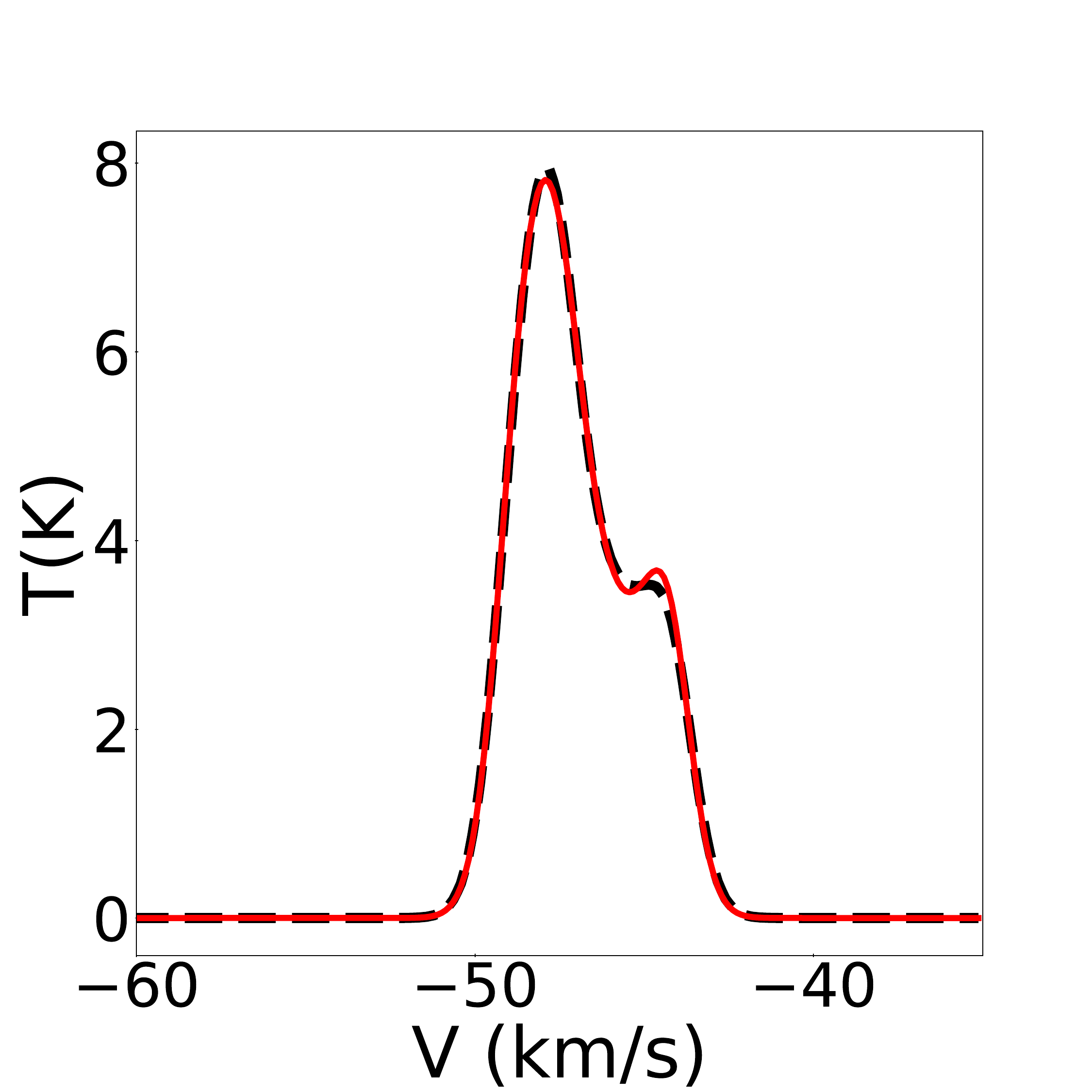}
   \includegraphics[width=0.30\columnwidth]{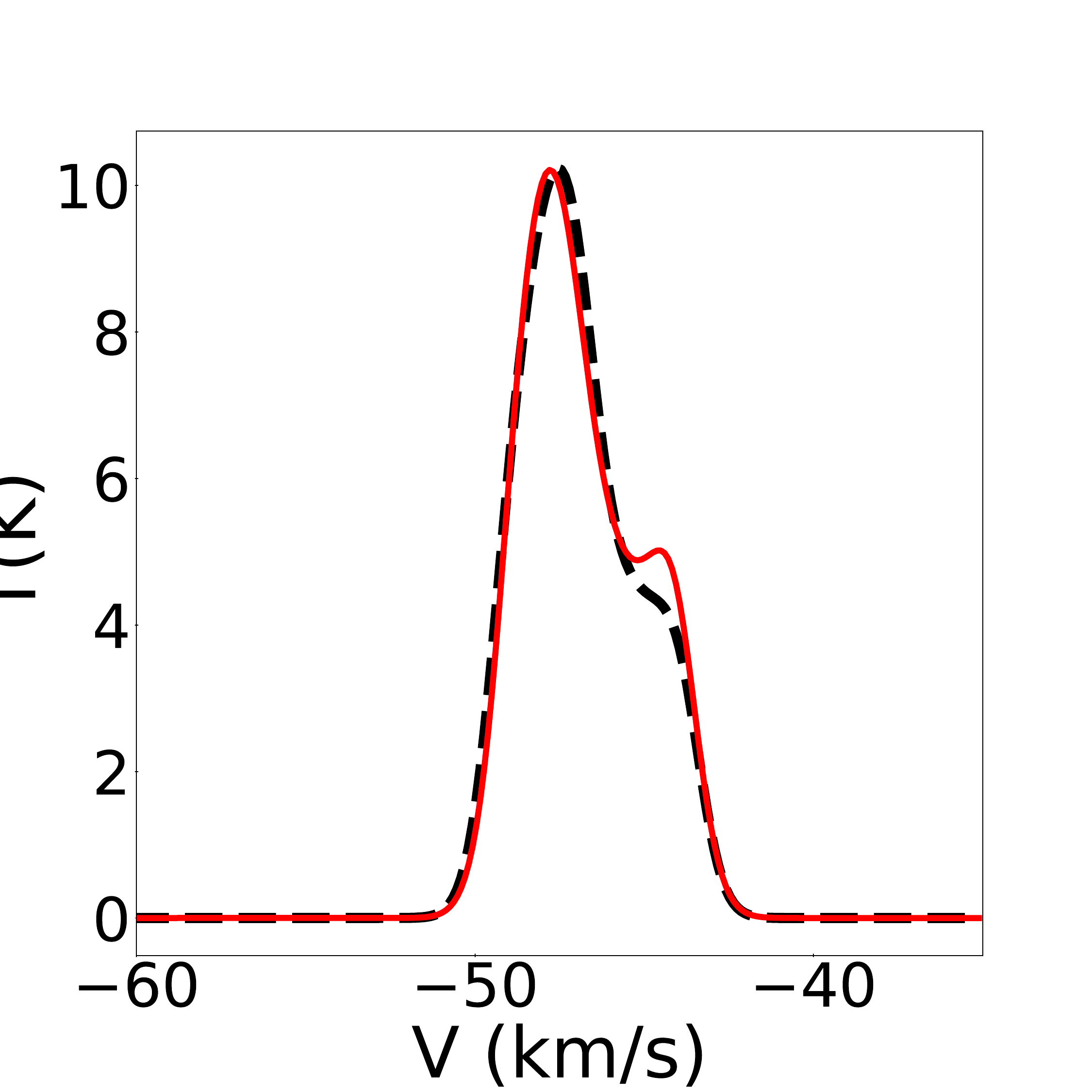}
\caption{Hill5 model fitting to the RT models. From left to right are RATRAN, LIME, and RADMC-3D results represented by the black dashed lines. The red profiles are the Hill5 fitting to the models. }
\label{fig: Hill_models}
\end{figure}

Applying Hill5 to the central pixel of the observations indicates an infall velocity of $-1.2$\,\kms, within 0.1\,\kms\ of the best fit uniform infall velocity models which give $-1.3$\,\kms for T=20 K ($M_{LIME,2}$ and $M_{LIME,7}$) and $-1.1$\,\kms for models with a temperature profile ($M_{LIME,1}$ and $M_{LIME,6}$). Hill5 fits a velocity dispersion of 1.05\,\kms\ identical to the 20 K models, but lower than the 1.13\,\kms\ of the temperature profile models, although this difference is likely not significant. 
Table~\ref{tab:Hill_models} and Figure~\ref{fig: Hill_models} show the results of applying Hill5 to the output of the uniform \vin\ radiative transfer models. Although Hill5 can fit the line profiles (Figure~\ref{fig: Hill_models}), it results in an underestimate of both the infall velocity and dispersion compared with the input values for all three models. This is consistent with \citet{2013ApJ...766..115K}, who argue that the results from Hill5 fitting should only be considered as a lower limit to the infall velocity.

\begin{table}
\caption{Weighted infall velocities for the best-fitting $p=2$ models with the most extreme values of $f$.}
\label{tab:weighted_vin}
\centering
\begin{tabular}{c c c c}
\hline
$f$ &
$v_c$ &
$\overline{v}_{\rm in,m}$ &
$\overline{v}_{\rm in,\rho}$ \\
&
($\mathrm{km\,s^{-1}}$) &
($\mathrm{km\,s^{-1}}$) &
($\mathrm{km\,s^{-1}}$) \\
\hline
$1/3$ & $-1.70$ & $-0.98$ & $-1.08$ \\
$3$   & $-0.60$ & $-0.87$ & $-0.38$ \\
\hline
\end{tabular}

\vspace{2pt}
\begin{minipage}{\columnwidth}
\footnotesize
\textit{Notes.}
$v_c$ is the infall velocity in the central region defined in equation~(\ref{eqn:vprofile});
$\overline{v}_{\rm in,m}$ is the mass-weighted infall velocity over the entire clump; and
$\overline{v}_{\rm in,\rho}$ is the density-weighted line-of-sight infall velocity towards the clump centre.
These values can be compared with the Hill5 fits to the central spectrum listed in Table~\ref{tab:bestFitHill5Models}.
\end{minipage}
\end{table}

This discrepancy can be larger when Hill5 is applied to models with non-uniform \vin. 
For models where the infall velocity varies with radius (Table~\ref{tab:bestFitModels}), depending on the form of this variation, Hill5 can either significantly underestimate or overestimate the infall velocity at centre of the clump. Towards the central position, Hill5 gives an infall velocity of $-1.2$\,\kms whereas models with central infall velocities as extreme as $-0.60$ and $-1.60$\,\kms can fit the observations (with different radial variations of the infall velocity). For comparison, Table~\ref{tab:weighted_vin} presents the mass- and density-weighted infall velocities for these extreme models, providing representative line-of-sight averaged velocities for both increasing and decreasing radial infall profiles. Comparing with the results in Table~\ref{tab:bestFitHill5Models} shows that Hill5 overestimates the infall velocity compared with these line-of-sight `averaged' values.  
Despite these shortcomings, in situations where models of the infall are not readily available, Hill5 provides a useful (and widely used) method to parameterise the infall velocity. However, as the exploration of the infall structure here shows, the results need to be interpreted with caution.

\subsection{Implications for massive star formation}    
 \label{subsec:implications} 
Different from low-mass star formation, where infall velocities are found to be around $-0.5$\,\kms \citep[e.g.][]{2006ApJ...637..860W,2013ApJ...766..115K}, massive star formation is found with higher infall velocities around $\sim-1.5$\,\kms \citep[e.g.][]{2012A&A...538A.140K, 2013A&A...549A...5R,2018ApJ...861...14C,2020ApJ...890...76N}. 
With a larger infall velocity, assuming a constant accretion throughout the formation, individual clump gains mass quicker in high-mass star-forming scenario than in lower-mass counterparts.

\begin{figure}
    \centering
    \includegraphics[width=0.4\textwidth]{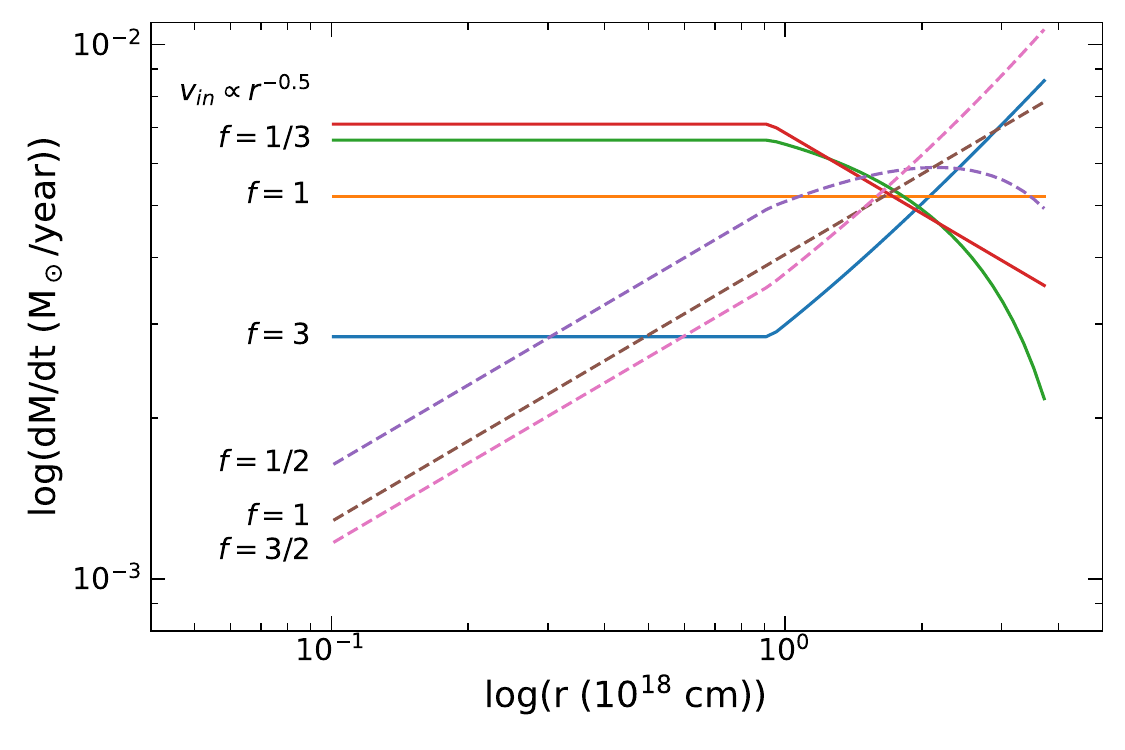}
    \caption{Mass accretion rates as a function of radius within the clumps for the best fit models of SDC335. The dashed curves show models with $p=1.5$ and the solid curves, models with $p=2$. The curves are labelled with the $f$ or the form of the infall velocity. 
%    {change f=0 to f=1. and make the figure nicer.}
}
    \label{fig:mdot_profiles}
    % Figure from BFProfilePlot.ipynb, using files made by PlotLIMEParameters.ipynb
\end{figure}

For massive star precursors, the accretion rates are greater than lower-mass counterparts, with the value from $\times$10$^{-4}$ to up to a few $\times$10$^{-3}$\,{\unitMsunyr} \citep[e.g.][]{2005A&A...442..949F, 2012A&A...538A.140K}. The previous calculation of infall velocity from \citet{2013A&A...555A.112P} estimated an accretion rate in SDC335 of (2.5$\pm$1.0)$\times$10$^{-3}$\,{\unitMsunyr} on the clump scale and \citet{2021A&A...645A.142A} have shown that the young outflows from the embedded protostars in SDC335 imply similar high values close to the protostars. Using an angular resolution of 0.3'' (corresponding to 1000 AU), \citet{2021ApJ...909..199O} used red-shifted absorption of $^{13}$CO\,J$=2-1$ to estimate an infall velocity of $-1.7$\,\kms with respect to a systemic velocity of $-46.9$\,\kms, which would correspond to an infall velocity of $-1.3$\,\kms with respect to the $-46.5$\,\kms\ adopted here. This is broadly consistent with recent ALMA results indicating that, on still smaller scales, the gas kinematics can reflect a transition between a rotating disc and a rotating, infalling envelope, with anisotropic accretion feeding the central protostellar system \citep{2026ApJ...999..106O}. Due to the low critical density and high optical depth of $^{13}$CO\,J$=2-1$, 
\citet{2021ApJ...909..199O} argued this traces the infall on a large size scale (rather than close to the central source) which would be consistent with the large infall velocities at large radius our results suggest. 
From the models here it is possible to describe the implied mass inflow rates as a function of radius in the clump. These are shown for the best fit models in Figure~\ref{fig:mdot_profiles}. The values are mostly in the range $10^{-3}$\,{\unitMsunyr} to $10^{-2}$\,{\unitMsunyr} with the spatial variation depending on the assumed density and infall velocity profile. Notice that the density profile has a significant impact on the range of the implied mass inflow rates. 

The inverted infall velocity structure, larger infall velocity at larger radii, implied by our three-dimensional RT modelling of SDC335 means that the mass flux has to be larger at larger radii. Such a configuration prevents any simple continuous mass flow toward the cloud centre. Infalling material will have to pile up or otherwise form structures at some intermediate scales, i.e. between the infall/cloud scales of $\sim$1 pc and the dense core scale of $\lesssim0.1\,\mathrm{pc}$. Thus, our model of the pc-scale infall suggest that fragmentation or the formation of filamentary structures will happen at intermediate scales for this massive star-forming region, as indeed is seen in higher resolution observations of this region \citep{2023MNRAS.520.3259X}.

\subsection{The origin of non-thermal motions}     

There has long been discussion of the origin of the supra-thermal line widths observed in molecular gas \citep[e.g.][]{1974ApJ...192L.149Z,1983ApJ...264..517M,2016A&A...591A.104H} and how their variation with size may trace the turbulent cascade of energy between size scales \citep[e.g.][]{2011MNRAS.411...65B,2019A&A...630A..97C,2019MNRAS.490.3061V}. More recently, \citet{2020MNRAS.491.4310T} suggested that the relatively shallow variation of observed line width with size when moving from their environment down to $\sim$pc-scale clumps, especially in high mass and high surface density regions, is a consequence of a contribution from infall to the line width and in SPH simulations of a collapsing clump by 
\cite{2007A&A...464..983P} the line broadening was dominated by the collapse.
Based on the relationship between velocity dispersion, size, and surface density for a sample of 70\,$\mu$m quiet clumps,
\citet{2018MNRAS.473.4975T} presented a similar interpretation. 
Using maps of a sample of IRDCs, \citet{2023MNRAS.525.2935P} have shown that 
the spatial profile of the velocity dispersion shows an abrupt flattening in slope when transitioning from clouds to their denser, parsec-scale, inner regions (clumps). After decreasing towards smaller size scales in the envelope, the velocity dispersion (as traced by N$_2$H$^+$\,J$=1-0$) is 
uniform across the clumps. 
This is interpreted as the pc-scale 
clumps being dynamically decoupled from their envelopes and suggesting the clumps are dominated by 
gravitational motions.

As Figure~\ref{fig:thin-dv} shows (Sec.~\ref{sec:h13cop_dv}), away from the central source, \HTCOp\,J$=1-0$ in SDC335 has a uniform distribution line width, similar to N$_2$H$^+$\,J$=1-0$ in the sources observed by \citet{2023MNRAS.525.2935P} and that infall can be responsible for this behaviour. 
Simulations suggest that the infall contribution to the observed velocity dispersion should be common in cores ($\sim0.1$\,pc-size regions) \citep[e.g.][]{2018MNRAS.479.2112B,2019A&A...625A..82N}
and supersonic line widths have been considered to trace the gravitational collapse where a virial-like relation developed from the gravity release \citep[e.g.][]{2011MNRAS.411...65B,2022MNRAS.515.2822R}, which is consistent with the enlarged line widths contributed from infall in our models.
Our models demonstrate the significant contribution of infall  
to the observed line widths of an optically thin dense gas tracer in massive clumps as was also seen in the simulations of \citet{2007A&A...464..983P}. This was also suggested by the ALMA core-scale study of \citet{2025ApJ...979..233M}. This opens the possibility of using the line widths of optically thin tracers to trace infall. 
%observational confirmation of its wider role in clumps, $\sim1$\,pc-sized regions, requires detailed analyses and modelling of the infall kinematics 
In addition, the previously poorly recognised contribution to the line width from infall can significantly affect the interpretation of the virial parameter, and hence the interpreted stability of clumps \citep[e.g.][]{2006MNRAS.372..443B,2018A&A...619L...7T,2018MNRAS.477.2220T,2021ApJ...922...87S}.

% GAFted from inGAF 
%\textfall in our modellings.color{red}{GAF: Maattribuny massive cores are apparently sub-virial. But this effect could make virial clumps look super-virial.  }

\section{Conclusions}\label{sec:concl}

We have carried out a systematic study of the infall in SDC335, a prototypical massive star-forming IRDC using the semi-analytical Hill5 model and radiative transfer models to study the emission of \HCOp. Focusing on models using LIME, we have explored the spectral line profiles produced by different infall velocity structures. Our main results are:

\begin{enumerate}

\item 
%The dominant factor in constraining the infall profiles is the assumed spatial structure of the infall velocity. 
A range of infall velocity structures can reproduce the blue asymmetric \HCOp\,J$=1-0$ line profile at the centre of the map with the best fit infall velocity in the central regions dependent on the assumed structure of the infall velocity in the outer regions of the clump.

\item
 In the central region of the clump SDC335 (at a radius of $<9.3\times10^{18}$\,cm, $<0.3$ pc), the magnitude of the infall velocities is  
constrained within a factor of $\sim3$ to be between 0.6 and 1.6\,\kms, with a weak dependence on the assumed cloud conditions, such as the density profile or temperature. 

\item
 The observed spectra show evidence of infall out to radii of $\sim6\times10^{18}$\,cm (1.95\,pc). Fitting the spectra with Hill5 shows a central peak in the inferred magnitude of the infall velocity of about 1.2\,\kms, followed by a decrease to about 0.6\,\kms at a radius $\sim2\times10^{18}$\,cm (0.65\,pc). At larger distances, Hill5 infers infall velocities $\sim-1$ \kms. 

\item
 Radiative transfer models show that the apparent decrease in the magnitude of the infall velocity inferred from Hill5 fits to \HCOp\ J=1--0 does not uniquely imply that the actual infall velocity is decreasing with increasing radius. For example, such a decrease can be reproduced with models in which the velocity of the infalling gas is greater at $2\times10^{18}$\,cm (0.65\,pc) than at the centre of the clump. Radiative transfer models show that the degeneracy between different structures for the infall velocity which can reproduce the structure seen in \HCOp\,J$=1-0$ could be broken with observations of additional \HCOp\ transitions. 

\item
Outside the central region, the line width of \HTCOp\,J$=1-0$ line is surprisingly constant. 
Comparison with synthetic spectra from our radiative transfer models show that this uniform line width has a significant contribution from unresolved infall motions over the entire clump. 
Typically, the line width in clumps is interpreted as due to turbulence,  
so this contribution from infall can significantly impact and alter our interpretation of the energetics and dynamical state of clumps. 

\item
Combining the comparison of the observed \HCOp\ and \HTCOp\ emission with the radiative transfer models favours an interpretation where the magnitude of the infall velocity is a maximum of $\sim1.5$ \kms\ towards centre of the clump and then decreases by a factor of 0.8 to about 1.2\,\kms\ at $\sim2\times10^{18}$\,cm ($\sim40''$) (Figure~\ref{fig:thin-dv}). Then at larger radii, the magnitude of the infall velocity is uniform or somewhat increasing with values of $\sim1.3$ to $1.6$\,\kms at a distance of $\sim4\times10^{18}$\,cm (1.30\,pc), consistent with the results of \citet{2021ApJ...909..199O}.

\item
The implied pc-scale mass inflow rates were found to be between $3\times10^{-3}$ and $10^{-2}$\,\Msun yr$^{-1}$. Clearly, the internal structure of SDC335 is in reality significantly more complex than the simple spherical models used here.
However, given the high optical depth and spatial extent of the \HCOp\,J$=1-0$ emission, the impact of such structure is difficult to assess without high angular resolution observations which sample the full range of size scales.  
Nevertheless, this first exploration of models applied to spatially extended infall signatures demonstrates the ability of modelling to constrain (the spherically averaged) infall properties in the region. 
\end{enumerate}
With the ever increasing volume of spatially-resolved infall signatures, modelling the line profiles on such spectral maps holds the promise to shed significant light on the dynamics of pc-scale infall.

\section*{Acknowledgements}

We want to thank the anonymous referee for his or her constructive suggestions, which have greatly improved this paper. This work is supported by the National Natural Science Foundation of China grant No. 12588202, the International Partnership Program of the Chinese Academy of Sciences grant No. 114A11KYSB20210010, the National Key R\&D Program of China No. 2022YFA1603103, Tianshan Talent Training Program 2024TSYCTD0013, and Natural Science Foundation of Xinjiang Uygur Autonomous Region No. 2025D01B173. It was also partially funded by the Regional Collaborative Innovation Project of Xinjiang Uygur Autonomous Region grant 2022E01050, Tianshan Talent Training Program, and Tianchi Talent Project of Xinjiang Uygur Autonomous Region.
X.J.J. acknowledges the support by the Chinese Scholarship Council (CSC) and the STFC China SKA Exchange Programme for support as a visiting PhD student in the United Kingdom. X. J. J. wants to thank Xu Dan and her Xiao Dan Gao. X. J. J. also wants to thank Dr. Qianru He and Xiale (Ruilin Xia) for the help on this paper.  
G.A.F acknowledges financial support from the State Agency for Research of the Spanish MCIU through the AYA 2017-84390-C2-1-R grant (co-funded by FEDER) and through the "Center of Excellence Severo Ochoa" award for the Instituto de Astrof\'isica de Andalucia
(SEV-2017-0709). 
G.A.F also acknowledges support from the Collaborative Research Centre 956, funded by the Deutsche Forschungsgemeinschaft (DFG) project ID 184018867. G.A.F also gratefully acknowledges the DFG for funding through SFB 1601 ``Habitats of massive stars across cosmic time'' (sub-project B1)  and from the University of Cologne and its Global Faculty programme. 
J. W thanks to the support of the Tianchi Talent Program of Xinjiang Uygur Autonomous Region.
This research made use of Astropy (\url{http://www.astropy.org}), a community-developed core Python package for Astronomy \citep[]{2013A&A...558A..33A, 2018AJ....156..123A,2022ApJ...935..167A} as well as the related packages Pyspeckit \citep{2022AJ....163..291G,2011ascl.soft09001G} and spectral-cube \citep{2016ginsburg,2019ginsburg}.

\noindent{Facilities: ATNF Mopra 22m}\\
{Software: RADMC-3D, RATRAN, LIME, idl, Astropy, Pyspeckit, spectral-cube}

\subsection*{Data Availability}

The data analysed in this paper are available from
\url{https://atoa.atnf.csiro.au/MALT90}.

%The Acknowledgements section is not numbered. Here you can thank helpful
%colleagues, acknowledge funding agencies, telescopes and facilities used etc.

%%%%%%%%%%%%%%%%%%%%%%%%%%%%%%%%%%%%%%%%%%%%%%%%%%



\bibliographystyle{mnras}
\bibliography{infall.bib} % if your bibtex file is called example.bib

%\end{thebibliography}

%%%%%%%%%%%%%%%%%%%%%%%%%%%%%%%%%%%%%%%%%%%%%%%%%%

%%%%%%%%%%%%%%%%% APPENDICES %%%%%%%%%%%%%%%%%%%%%

\appendix

\section{Line Characterisation}\label{sec:linechara}

%We describe here a method to characterise or parameterise infall line profiles. 
To evaluate the fitting of the modelling results to the observed profiles, we develop a method to parameterise the line profiles of infall features. There are four parameters that can be used to encapsulate the characteristic blue asymmetric line profile:

\begin{enumerate}
    \item $P_r=T_{\rm blue}/T_{\rm red}$, the ratio of the peak temperature of the blue-shifted peak to the peak temperature of the red-shifted peak, 
    \item  $v_{\rm sep} = v_{\rm red}-v_{\rm blue}$, the velocity separation of the two peaks, 
    \item $w_{1/2}$, the full width of the line profile at half the peak intensity of the red side of the profile, which guarantees both the red and blue sides of the line are encompassed, and
    \item $D_{\rm r} = T_{\rm dip}/T_{\rm blue}$, the ratio of the intensity at the minimum between the two peaks to the peak intensity of the blue-shifted peak. 
\end{enumerate}

\begin{figure} 
   \centering
   \includegraphics[width=\columnwidth]{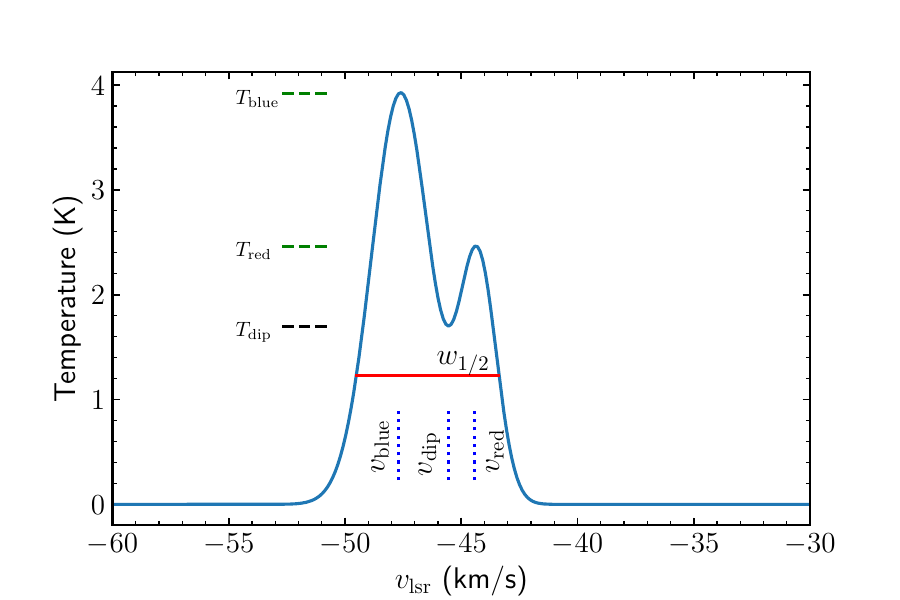}
\caption{The characterisation of the line profile to quantify infall signatures. T$_{blue}$ is the peak temperature of the blue side of the profile, while T$_{red}$ is that of the red side. T$_{dip}$ is the temperature at the absorption dip. v$_{blue}$ and v$_{red}$ are the corresponding velocities at the blue and red peaks, respectively. v$_{dip}$ is the corresponding velocity at the absorption dip. $w_{1/2}$ is the full width of the line profile at half the peak intensity of the red side of the profile. }
\label{fig: characterisation}
% Figure from 
\end{figure}

Figure~\ref{fig: characterisation} displays these parameters on a typical infall profile. For our analysis, we selected the pixel in the observed data with the largest values of the parameter $P_r$. This pixel corresponds to the position ($l$,$b$)$=335.5836^\circ, -0.2862^\circ$. 
%({change this.}RA(J2000),Dec(J2000)=  hms,dms). 
% Note: This is pixel 13,13 in the maps!! 19 July 2022
The parameter values and their uncertainties measured for the observed central pixel are given in Table~\ref{tab:observed_param}.

\begin{table}
    \centering
        \caption{Measured line characterization parameters and their uncertainty for the central pixel of the observations. }
%    Vsep = 3284.45+/-113.26
%Width = 6455.63+/-113.26
% from velocity -49737.82 to -43282.19
%PeakR = 2.09+/-0.29
%DipRatio = 0.39+/-0.06
\label{tab:observed_param}
    \begin{tabular}{cccc}
    \hline
   $P_r=T_{\rm blue}/T_{\rm red}$ & $v_{\rm sep} = v_{\rm red}-v_{\rm blue}$ & $w_{1/2}$ & $D_{\rm r} = T_{\rm dip}/T_{\rm blue}$\\
    & & (km/s) & (km/s)\\
    \hline
     2.1 (0.3) & 3.3 (0.1) & 6.4 (0.1) & 0.39 (0.06)  \\
    \hline
    \end{tabular}
\end{table}

Figure~\ref{fig:infallParamMaps} 
shows the spatial distribution of $P_r$,  the ratio of the intensity of the blue-shifted peak to that of the red-shifted peak, across the SDC335. The map shows that the spectroscopic signature of infall is seen over a significant portion of the clump.  
%While providing a description of the line shape these parameters do not  on their own provide any estimate of the properties of the infall. To do this requires a model. The simplest and most widely used are models are based in a simple layer approximation \citep{XXX}, with the most sophisticated of  these being  the Hill model \citep{XXX}.

%\begin{figure} 
%   \centering
%   \includegraphics[width=\columnwidth]{r-map.pdf}
%\caption{The spatial distribution of the $P_r$, the intensity ratio of the blue-shifted peak to the red-shifted peak. The parameters are evaluated for the region where the integrated intensity of $>20$\% of the peak integrated intensity in the line (as indicted by the contour). The red cross corresponds to the highest value of $P_{r}$. \
%textcolor{red}{make this consistent with Fig. 15. } - Done
%}
%\label{fig:infallParamMaps}
% Figure from notebook: AnalysisNotebooks/InfallFullCube.ipynb
%\end{figure}


%%%%%%%%%%%%%%%%%%%%%%%%%%%%%%%%%%%%%%%%%%%%%%%%%%

\section{Model $v_{\rm c}-$\dopb\ Planes}

\begin{landscape}
\begin{figure}
    \centering
    \includegraphics[width=\linewidth]{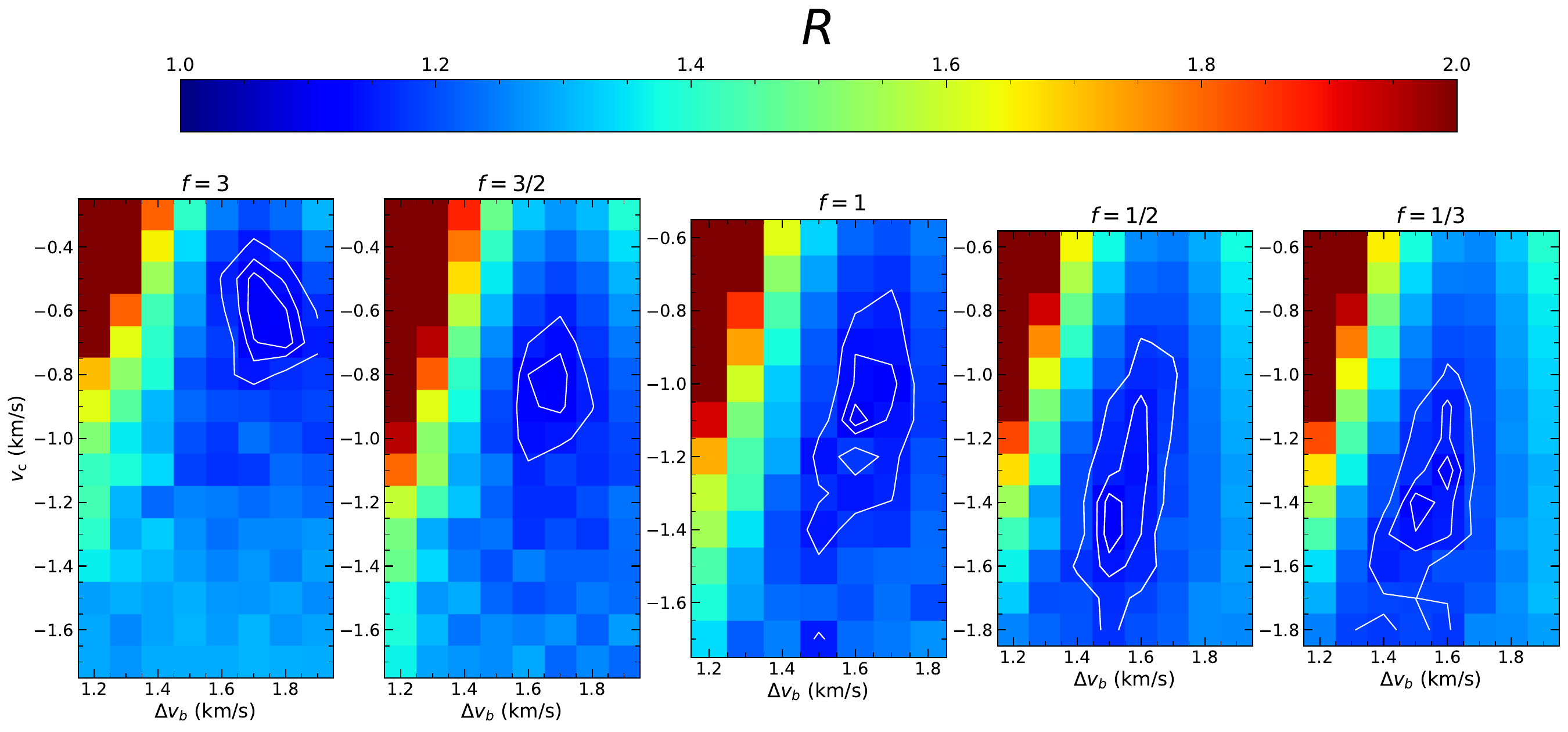}
    \caption{Colour scale images of the $R$ parameter in  $v_{\rm c}-$\dopb plane for the central pixel of the observations. The axes show the full range of 
$v_{\rm in,c}$ and \dopb\ explored. Both values were sampled every 0.1\,\kms. The panels show the $p=2$ models with the $f$ as labelled at the top of each panel. From left to right these are models $M_{LIME, 10}$, $M_{LIME, 9}$, $M_{LIME, 6}$, $M_{LIME, 5}$ and $M_{LIME, 4}$. The contours show 1.01, 1.025 and 1.05 times the minimum $R$ for the grid of models. From left to right this minimum value is 1.13, 1.12, 1.11, 1.10, 1.11.}
    \label{fig:other_vin_db-plane}
\end{figure}

\end{landscape}

Figure~\ref{fig:other_vin_db-plane} shows the $v_{\rm c}-$\dopb\ planes for the $p=2$ LIME models (apart from model  $M_{LIME,7}$ which is shown in Fig.~\ref{fig:vin_db-plane} in the main text). 

% Don't change these lines
\bsp	% typesetting comment
\label{lastpage}
\end{document}